\documentclass[twocolumn]{aastex631}

\usepackage{graphicx}
\usepackage[varg]{txfonts}
\usepackage{cases}
\usepackage{natbib}
\usepackage{multirow}
\usepackage{longtable}
\usepackage{graphicx}
\usepackage{booktabs}

\shorttitle{Statistical analysis of circular-ribbon flares}
\shortauthors{Zhang et al.}

\graphicspath{{./}{figures/}}

\begin{document}

\title{Statistical analysis of circular-ribbon flares}


\author[0000-0003-1979-9863]{Yanjie Zhang}
\affiliation{Key Laboratory of Dark Matter and Space Astronomy, Purple Mountain Observatory, CAS, Nanjing 210023, China}
\affiliation{School of Astronomy and Space Science, University of Science and Technology of China, Hefei 230026, China}

\author[0000-0003-4078-2265]{Qingmin Zhang}
\affiliation{Key Laboratory of Dark Matter and Space Astronomy, Purple Mountain Observatory, CAS, Nanjing 210023, China}
\affiliation{School of Astronomy and Space Science, University of Science and Technology of China, Hefei 230026, China}

\author[0000-0003-0057-6766]{Dechao Song}
\affiliation{Key Laboratory of Dark Matter and Space Astronomy, Purple Mountain Observatory, CAS, Nanjing 210023, China}
\affiliation{School of Astronomy and Space Science, University of Science and Technology of China, Hefei 230026, China}

\author[0000-0003-2694-2875]{Shuting Li}
\affiliation{Key Laboratory of Dark Matter and Space Astronomy, Purple Mountain Observatory, CAS, Nanjing 210023, China}
\affiliation{School of Astronomy and Space Science, University of Science and Technology of China, Hefei 230026, China}

\author[0000-0002-0786-7307]{Jun Dai}
\affiliation{Key Laboratory of Dark Matter and Space Astronomy, Purple Mountain Observatory, CAS, Nanjing 210023, China}
\affiliation{School of Astronomy and Space Science, University of Science and Technology of China, Hefei 230026, China}

\author[0000-0002-9121-9686]{Zhe Xu}
\affiliation{Yunnan Observatories, Chinese Academy of Sciences, 396 Yangfangwang, Guandu District, Kunming 650216, China}

\author[0000-0002-5898-2284]{Haisheng Ji}
\affiliation{Key Laboratory of Dark Matter and Space Astronomy, Purple Mountain Observatory, CAS, Nanjing 210023, China}
\affiliation{School of Astronomy and Space Science, University of Science and Technology of China, Hefei 230026, China}

\begin{abstract}
Circular-ribbon flares (CFs) are a special type of solar flares owing to their particular magnetic topology.
In this paper, we conducted a comprehensive statistical analysis of 134 CFs from 2011 September to 2017 June, 
including four B-class, 82 C-class, 40 M-class, and eight X-class flares, respectively.
The flares were observed by the Atmospheric Imaging Assembly (AIA) on board the Solar Dynamics Observatory (SDO) spacecraft.
The physical properties of CFs are derived, including the location, area ($A_{CF}$), equivalent radius ($r_{CF}$) assuming a semi-spherical fan dome, 
lifetime ($\tau_{CF}$), and peak SXR flux in 1$-$8 {\AA}.
It is found that all CFs are located in active regions, with the latitudes between -30$\degr$ and 30$\degr$.
The distributions of areas and lifetimes could be fitted with a log-normal function. There is a positive correlation between the lifetime and area.
The peak SXR flux in 1$-$8 {\AA} is well in accord with a power-law distribution with an index of $-$1.42.
For the 134 CFs, 57\% of them are accompanied by remote brightenings or ribbons.
A positive correlation exists between the total length ($L_{RB}$) and average distance ($D_{RB}$) of remote brightenings.
About 47\% and 51\% of the 134 CFs are related to type III radio bursts and jets, respectively. The association rates are independent of flare energies.
About 38\% of CFs are related to mini-filament eruptions, and the association rates increase with flare classes.
Only 28\% of CFs are related to CMEs, meaning that a majority of them are confined rather than eruptive events.
There is a positive correlation between the CME speed and peak SXR flux in 1$-$8 {\AA}, and faster CMEs tend to be wider.
\end{abstract}


\section{Introduction} \label{intro}
Since the first discovery in 1859 \citep{Car1859}, solar flares have been observed and studied extensively \citep[see][and references therein]{Fle2011,Shi2011}. 
According to the classical two-dimensional (2D) flare model, i.e., the CSHKP model \citep{Car1964,Stu1966,Hir1974,Kop1976}, when high-energy electrons accelerated by magnetic reconnection 
propagate downward along the reconnected flare loops and hit the chromosphere, the localized plasmas at the footpoints are impulsively heated, forming two bright and elongated ribbons 
observed in ultraviolet (UV), extreme-ultraviolet (EUV), and H$\alpha$ wavelengths \citep{ji06,Jing2016}. In most cases, the ribbons separate due to continuing reconnection \citep{qiu02}.
\citet{Aul2012} and \citet{Jan2013} extended the 2D standard model to three dimensions to interpret the flare ribbons taking on double-J shapes.

Apart from the typical two-ribbon flares, there is another class of flares, i.e., circular-ribbon flares (CFs), whose ribbons are elliptical or circular \citep{Mas2009}.
It is generally believed that the three-dimensional (3D) magnetic configuration of CFs is composed of a null point and the associated dome-shaped fan-spine structure in the corona 
\citep[e.g.,][]{Rei2012,Wang2012,Jia2013,Sun2013,VW2014,Jos2015,Jos2021,Liu2015,Zha2015,Zha2021,Mi2021}. 
Sometimes, the outer spine is embedded in a thin quasi-separatrix layer \citep[QSL;][]{Dem1996} where magnetic connectivity changes rapidly \citep{Mas2009,Yang2015,Li2018}.
Magnetic reconnection preferentially occurs near the null point \citep{Pri2009,Pon2007,Pon2013,Yang2020b}, 
and the closed ribbon is related to the intersection between the chromosphere and fan surface.
Meanwhile, a shorter ribbon inside the closed ribbon is believed to be the intersection between the chromosphere and inner spine.
The sizes of circular ribbons range from $\sim$30$\arcsec$ \citep{Wang2012,Hao2017} to several 100$\arcsec$ \citep{Liu2013,Jos2017,Hou2019,Chen2019,Lee2020a}.
The brightening of circular flare ribbons is not always simultaneous, but sequential in some cases, which is probably caused by slip-running reconnection \citep{Aul2007}.
The direction of brightening could be clockwise \citep{Shen2019} or counterclockwise \citep{Mas2009,Li2017,Li2018,Xu2017}.
Interestingly, \citet{Zha2020} reported fast degradation of a circular ribbon at speeds of 30$-$70 km s$^{-1}$ on 2014 August 24.
Remote brightenings are frequently observed adjacent to the main flare site as a result of energy transport along the outer spine \citep{Mas2009,Mas2017,Zha2015,Her2017,Li2017,Devi2020,Jos2021}. 
Extended remote brightenings with a length of $\sim$400$\arcsec$ is investigated by \citet{Liu2020}.

CFs are usually triggered by filament eruptions \citep{Jos2015,Xu2017,Song2018,Yang2020a}, which may also generate coronal jets \citep{Wang2012,Jos2018,LY2019,Zha2016a,Zha2021,ZN2019,ZQM2020,Dai2021}
or drive coronal mass ejections \citep[CMEs;][]{Sun2013,Jos2015,Jos2017,Liu2019,Liu2020}. CFs show similar dynamics to that of two-ribbon flares. 
Explosive chromospheric evaporation is detected in a C4.2 class flare by the Interface Region Imaging Spectrograph \citep[IRIS;][]{De2014} on 2015 October 16 \citep{Zha2016a}.
The estimated energy flux of nonthermal electrons is adequate to account for the explosive evaporation. Imaging observation of converging hot ($\log T\approx7.0$) 
plasma in a post flare loop is considered as direct evidence of chromospheric evaporation during the impulsive phase of a flare \citep{Zha2019a}.
Redshifts of a few 10 km s$^{-1}$ in the low-temperature emission lines (e.g., Si {\sc iv}) are clear indications of chromospheric condensation at the circular flare ribbons \citep{Zha2016b,Zha2021}.
Besides, quasi-periodic pulsations \citep[QPPs;][]{Zim2021} produced by intermittent magnetic reconnections are identified in CFs \citep{Kum2015,Kum2016,Zha2016b,Chen2019,Kash2020,Lee2020b,Li2020,Ning2022}. The periods are between 20 s and 4 minutes in most cases. 
CFs are also accompanied by type III radio bursts \citep{Zha2016b,Zha2021} or type IV radio continuum emission \citep{Chen2019}.
The total energies of CFs can reach up to 10$^{31}$-10$^{33}$ erg. \citet{Zha2019b} calculated various energy components in two successive M1.8 class CFs in NOAA active region (AR) 12434, 
including the peak thermal energy, nonthermal energy in flare-accelerated 
electrons, total radiative loss of the hot plasma, and radiative energy in 1$-$8 {\AA} and 1$-$70 {\AA}. It is revealed that the energy partitions in two flares are similar, 
and the heating requirement consisting of the peak thermal energy and radiative loss could sufficiently be supplied by the nonthermal energy.
Further more, \citet{Cai2021} investigated four confined CFs in detail, finding that the values of energy components increase systematically with flare classes.
The ratio of nonthermal energy to magnetic free energy may provide a key factor for discriminating confined from eruptive flares.

So far, CFs have become a topic of great interest due to their particular configurations. However, statistical investigation of CFs is rare compared to two-ribbon 
flares \citep[e.g.,][]{Cro1993,Tem2001,Ver2002,Yas2006,Li2021,Lu2021} or microflares \citep{Chr2008,Han2008}.
\citet{ST2018} performed a statistical investigation of 90 CFs observed from 2010 June to 2017 December, finding that the occurrence rate of white-light (WL) flares increases with flare class. 
Moreover, the flares with WL enhancement have shorter durations, smaller sizes, stronger electric current, and more complicated magnetic field.
Nevertheless, a comprehensive study of physical properties of CFs and related phenomena is scarce.
In this paper, we carry out a statistical analysis of 134 CFs, including four B-class, 82 C-class, 40 M-class and eight X-class CFs (see Table~\ref{tab:list}).
The paper is organized as follows. We describe the data analysis in Section~\ref{data}. Statistical properties are presented in Section~\ref{sta}.
Finally, a brief conclusion and discussions are given in Section~\ref{sum}.

\section{Data analysis} \label{data}
We searched for CFs observed by the Atmospheric Imaging Assembly \citep[AIA;][]{Lem2012} on board the Solar Dynamics Observatory \citep[SDO;][]{Pes2012} from 2011 September to 2017 June.
AIA took full-disk images in seven EUV (94, 131, 171, 193, 211, 304, and 335 {\AA}) and two UV (1600 and 1700 {\AA}) wavelengths.
The AIA level\_1 data with a time cadence of 12 s and a spatial resolution of 1$\farcs$2 were calibrated using the standard program \texttt{aia\_prep.pro} in the Solar Software (SSW).
Photospheric line-of-sight (LOS) magnetograms of the flares were observed by the Helioseismic and Magnetic Imager \citep[HMI;][]{sch12} on board SDO.
The HMI level\_1 data with a time cadence of 45 s and a spatial resolution of 1$\farcs$2 were calibrated using the standard program \texttt{hmi\_prep.pro}.
Soft X-ray (SXR) light curves of CFs in 1$-$8 {\AA} were recorded by the GOES spacecraft.
The associations with remote brighenings, coronal jets, and mini-filament eruptions were checked using the EUV observations of AIA.
The total lengths of remote brightenings and average distances from the flare centers were calculated.
The association with CMEs was examined using the WL observations from the Large Angle and Spectrometric Coronagraph \citep[LASCO;][]{Bru1995} on board the SOHO spacecraft.
The LASCO/C2 has a field of view (FOV) of 2$-$6 $R_\sun$.
Two databases were used: the CDAW catalog $\footnote{http://cdaw.gsfc.nasa.gov/CME\_list}$ where CMEs are identified manually and 
the CACTus website $\footnote{http://sidc.oma.be/cactus/scan}$ where CMEs are recognized automatically \citep{Yas2008}.
The apparent linear velocities and angular widths of the associated CMEs were analyzed.
The relation with type III radio bursts was checked using the radio dynamic spectra recorded by the e-Callisto\footnote{http://www.e-callisto.org} ground-based stations 
as well as the WAVES \citep{Bou1995} instrument (0.02$-$13.825 MHz) on board the WIND spacecraft \footnote{https://solar-radio.gsfc.nasa.gov/wind/index.html}.

\setlength{\tabcolsep}{4pt}
\begin{center}
\begin{longtable*}{cccccccccccccc}
	\caption{\centering List of 134 Circular-ribbon Flares.}
	\label{tab:list} \\
	\hline
No. & NOAA & Date & GOES   & Peak Flux & Location & $\tau_{CF}$ & $A_{CF}$  & $L_{RB}/D_{RB}$ & Jet & Type \textrm{III} & FE  & $V_{CME}$    & $W_{CME}$     \\ 
 &  AR   &         & Class  &   (W m$^{-2}$)  &      &  (min)      & (Mm$^{2}$)&     (Mm/Mm)          &     &                     &    &(km s$^{-1}$) & ($^\circ$)  \\
	\hline
	\endhead
1 & 11283 & 2011 Sep 06 & M5.3 & 5.39E-05 & N15W09 & 94 & 2823.5 & 28.0/117.6 & No & No & No  & 782 & 360 \\ 
2 & 11283 & 2011 Sep 06 & X2.1 & 2.16E-04 & N15W20 & 35 & 2594.8 & 65.8/105.6 & No & No & Yes & 575 & 360 \\
3 & 11283 & 2011 Sep 07 & X1.8 & 1.80E-04 & N16W32 & 81 & 1983.0 & 193.4/76.3 & No & No & Yes & 792 & 167 \\
4 & 11283 & 2011 Sep 08 & M6.7 & 6.77E-05 & N16W42 & 64 & 2439.6 & 103.0/77.9 & No & No & Yes & 983 & 281 \\
5 & 11324 & 2011 Oct 22 & C4.1 & 4.12E-06 & N11E24 & 34 & 1616.9 & 19.3/63.7  & No & No & No  & -   & -   \\
6 & 11339 & 2011 Nov 03 & X1.9 & 2.04E-04 & N21E64 & 102& 725.8  & 65.2/35.7  & No & No & Yes & 991 & 360 \\
7 & 11339 & 2011 Nov 06 & M1.4 & 1.45E-05 & N21E32 & 94 & 581.7  & 14.1/70.7  & No & Yes& No  & -   & -   \\
8 & 11339 & 2011 Nov 06 & C8.8 & 9.10E-06 & N21E30 & 43 & 542.6  & 35.7/67.5  & No & No & No  & -   & -   \\
9 & 11339 & 2011 Nov 06 & C5.3 & 5.42E-06 & N21E28 & 113& 422.8  & 33.2/72.2  & No & No & No  & -   & -   \\
10 & 11346 & 2011 Nov 15 & M1.9 & 1.97E-05 & S19E32 & 99 & 1165.3 & 144.8/86.1 & Yes& Yes& No  & 163 & 80  \\
11 & 11346 & 2011 Nov 16 & C2.8 & 2.91E-06 & S18E19 & 31 & 315.8  & 91.3/75.2  & Yes& No & Yes & -   & -   \\
12 & 11346 & 2011 Nov 16 & C2.9 & 3.01E-06 & S18E16 & 37 & 345.1  & 59.8/72.7  & Yes& No & Yes & -   & -   \\
13 & 11346 & 2011 Nov 16 & C5.0 & 5.14E-06 & S19E12 & 59 & 1181.3 & 121.8/118.3& Yes& No & Yes & -   & -   \\
14 & 11346 & 2011 Nov 17 & C2.0 & 2.05E-06 & S19E04 & 59 & 1278.9 & 134.1/86.7 & No & No & No  & -   & -   \\
15 & 11476 & 2012 May 06 & C1.4 & 1.43E-06 & N11E72 & 12 & 354.0  &    -       & No & No & No  & -   & -   \\
16 & 11476 & 2012 May 07 & C4.0 & 4.17E-06 & N11E61 & 73 & 598.9  & 39.6/54.9  & Yes& Yes& No  & -   & -   \\
17 & 11476 & 2012 May 07 & C7.9 & 8.07E-06 & N13E60 & 24 & 428.0  & 78.3/67.7  & No & Yes& No  & -   & -   \\
18 & 11476 & 2012 May 07 & C7.4 & 7.67E-06 & N13E56 & 59 & 445.1  &    -       & No & Yes& No  & -   & -   \\
19 & 11476 & 2012 May 08 & M1.4 & 1.48E-05 & N13E45 & 92 & 545.6  & 60.1/92.3  & No & Yes& Yes & -   & -   \\
20 & 11476 & 2012 May 09 & M4.7 & 4.86E-05 & N13E32 & 61 & 569.4  & 79.2/91.7  & Yes& No & Yes & -   & -   \\
21 & 11476 & 2012 May 09 & C1.5 & 1.55E-06 & N13E28 & 11 & 424.1  &    -       & Yes& No & No  & -   & -   \\
22 & 11476 & 2012 May 09 & M4.1 & 4.16E-05 & N13E27 & 66 & 797.7  & 128.2/97.4 & Yes& No & Yes & -   & -   \\
23 & 11476 & 2012 May 10 & M5.7 & 6.00E-05 & N13E23 & 101& 1188.3 & 119.5/86.2 & Yes& Yes& Yes & -   & -   \\
24 & 11476 & 2012 May 10 & M1.7 & 1.84E-05 & N13E14 & 83 & 979.8  & 82.5/98.5  & Yes& No & Yes & -   & -   \\
25 & 11476 & 2012 May 11 & C3.1 & 3.15E-06 & N14E13 & 20 & 816.2  &    -       & Yes& Yes& No  & -   & -   \\
26 & 11476 & 2012 May 11 & C3.5 & 3.69E-06 & N14E06 & 11 & 485.9  &    -       & Yes& Yes& No  & -   & -   \\
27 & 11476 & 2012 May 14 & C2.5 & 3.02E-06 & N08W46 & 36 & 339.8  & 27.4/41.3  & Yes& Yes& No  & 551 & 48  \\
28 & 11598 & 2012 Oct 22 & M5.0 & 5.10E-05 & S13E64 & 113& 1343.4 &    -       & No & No & No  & -   & -   \\
29 & 11598 & 2012 Oct 23 & X1.8 & 1.71E-04 & S13E59 & 108& 1423.2 &    -       & No & No & No  & -   & -   \\
30 & 11652 & 2013 Jan 08 & C1.8 & 1.92E-06 & N21E56 & 25 & 343.5  &    -       & Yes& No & No  & -   & -   \\
31 & 11652 & 2013 Jan 12 & C3.1 & 3.29E-06 & N19W17 & 4  & 182.1  &    -       & No & No & No  & -   & -   \\
32 & 11652 & 2013 Jan 13 & M1.0 & 1.25E-05 & N18W18 & 16 & 311.1  &    -       & Yes& Yes& No  & -   & -   \\
33 & 11652 & 2013 Jan 13 & C2.7 & 2.83E-06 & N18W22 & 20 & 213.6  &    -       & No & Yes& No  & -   & -   \\
34 & 11652 & 2013 Jan 13 & M1.7 & 1.86E-05 & N18W22 &  9 & 362.1  & 56.2/115.2 & Yes& Yes& Yes & 696 & 46  \\
35 & 11652 & 2013 Jan 14 & C6.5 & 6.75E-06 & N18W31 & 14 & 438.4  & 24.2/66.6  & No & No & No  & -   & -   \\
36 & 11669 & 2013 Feb 05 & B6.6 & 6.92E-07 & N09E64 & 63 & 1191.6 &    -       & Yes& No & Yes & -   & -   \\
37 & 11669 & 2013 Feb 05 & C6.3 & 6.51E-06 & N08E62 & 25 & 1649.0 &    -       & Yes& Yes& Yes & 444 & 66  \\
38 & 11675 & 2013 Feb 17 & M1.9 & 2.80E-05 & N12E21 & 10 & 183.0  & 7.0/47.7   & Yes& Yes& No  & -   & -   \\
39 & 11689 & 2013 Mar 12 & C3.6 & 3.87E-06 & S20W40 & 28 & 154.2  & 21.3/82.8  & Yes& No & Yes & -   & -   \\
40 & 11731 & 2013 Apr 28 & C3.7 & 4.07E-06 & N09E28 & 9  & 335.2  &    -       & Yes& Yes& No  & -   & -   \\
41 & 11731 & 2013 Apr 28 & C1.8 & 1.96E-06 & N09E28 & 8  & 433.3  &    -       & Yes& No & Yes & -   & -   \\
42 & 11731 & 2013 Apr 28 & C3.6 & 4.02E-06 & N09E27 & 13 & 133.2  &    -       & Yes& Yes& No  & -   & -   \\
43 & 11731 & 2013 Apr 29 & C3.0 & 3.06E-06 & N10E14 & 14 & 634.7  &    -       & No & Yes& No  & 452 & 67  \\
44 & 11731 & 2013 May 01 & C1.1 & 1.14E-06 & N07W11 & 10 & 160.8  &    -       & Yes& No & Yes & -   & -   \\
45 & 11731 & 2013 May 02 & M1.1 & 1.13E-05 & N11W25 & 43 & 4539.9 &    -       & Yes& Yes& Yes & -   & -   \\
46 & 11890 & 2013 Nov 05 & M2.5 & 2.81E-05 & S16E51 & 12 & 752.7  & 135.3/72.1 & No & Yes& Yes & -   & -   \\
47 & 11890 & 2013 Nov 05 & X3.3 & 3.85E-04 & S13E45 & 23 & 796.4  & 61.8/79.4  & Yes& Yes& Yes & 562 & 195 \\
48 & 11890 & 2013 Nov 06 & C8.6 & 8.93E-06 & S13E39 & 17 & 726.7  &    -       & Yes& Yes& Yes & -   & -   \\
49 & 11890 & 2013 Nov 06 & M3.8 & 3.88E-05 & S13E36 & 21 & 701.0  & 169.8/149.8& No & Yes& Yes & 347 & 122 \\
50 & 11890 & 2013 Nov 07 & M2.3 & 2.41E-05 & S13E28 & 22 & 727.8  &    -       & Yes& No & Yes & -   & -   \\
51 & 11890 & 2013 Nov 07 & C5.9 & 5.99E-06 & S13E23 & 21 & 697.8  &    -       & No & Yes& Yes & -   & -   \\
52 & 11890 & 2013 Nov 08 & X1.1 & 1.22E-04 & S14E15 & 24 & 1485.7 & 311.1/120.9& Yes& Yes& Yes & -   & -   \\
53 & 11890 & 2013 Nov 10 & X1.1 & 1.14E-04 & S15W13 & 33 & 2171.1 & 381.0/119.1& Yes& Yes& Yes & 682 & 262 \\
54 & 11890 & 2013 Nov 10 & C3.1 & 3.17E-06 & S15W19 & 56 & 306.5  &    -       & Yes& Yes& No  & -   & -   \\
55 & 11890 & 2013 Nov 11 & C6.4 & 6.59E-06 & S15W26 & 33 & 521.0  & 39.6/110.1 & No & No & Yes & -   & -   \\
56 & 11890 & 2013 Nov 11 & C7.8 & 8.18E-06 & S14W24 & 12 & 642.7  & 12.5/128.0 & Yes& No & No  & 533 & 40  \\
57 & 11890 & 2013 Nov 11 & C5.0 & 5.09E-06 & S14W34 & 25 & 1020.6 & 16.7/105.6 & Yes& No & No  & -   & -   \\
58 & 11890 & 2013 Nov 12 & C3.1 & 3.22E-06 & S11W60 & 12 & 346.8  &    -       & Yes& No & No  & 365 & 24  \\
59 & 11890 & 2013 Nov 13 & C6.5 & 6.75E-06 & S14W54 & 144& 877.3  & 50.2/84.0  & Yes& Yes& No  & -   & -   \\
60 & 11936 & 2013 Dec 25 & C1.7 & 1.79E-06 & S18E51 & 58 & 621.2  &    -       & No & No & No  & 254 & 72  \\
61 & 11936 & 2013 Dec 27 & C4.4 & 4.52E-06 & S17E23 & 35 & 507.7  &    -       & Yes& Yes& No  & 305 & 67  \\
62 & 11936 & 2013 Dec 28 & C3.0 & 3.08E-06 & S17E09 & 38 & 931.1  & 32.3/49.2  & No & Yes& Yes & -   & -   \\
63 & 11936 & 2013 Dec 28 & C9.3 & 9.44E-06 & S17E06 & 100& 684.1  & 24.5/45.0  & No & No & No  & -   & -   \\
64 & 11936 & 2013 Dec 29 & C3.1 & 3.17E-06 & S17W01 & 9  & 451.6  &    -       & No & Yes& No  & -   & -   \\
65 & 11936 & 2013 Dec 29 & M3.1 & 3.23E-05 & S17W02 & 65 & 828.1  & 10.4/48.6  & Yes& Yes& Yes & -   & -   \\
66 & 11936 & 2013 Dec 29 & C5.1 & 5.26E-06 & S16W06 & 60 & 831.7  & 7.7/48.9   & No & No & No  & -   & -   \\
67 & 11936 & 2013 Dec 29 & C1.9 & 2.02E-06 & S16W07 & 20 & 364.2  &    -       & No & No & No  & -   & -   \\
68 & 11936 & 2013 Dec 29 & C5.4 & 5.57E-06 & S16W09 & 61 & 647.6  & 218.6/103.0& Yes& No & No  & -   & -   \\
69 & 11936 & 2013 Dec 31 & M6.4 & 6.49E-05 & S17W36 & 128& 2582.3 & 240.5/88.9 & No & Yes& Yes & -   & -   \\
70 & 11936 & 2014 Jan 01 & M9.9 & 1.00E-04 & S16W47 & 55 & 4240.8 & 116.8/84.2 & No & No & Yes & 326 & 113 \\
71 & 11991 & 2014 Feb 27 & C6.8 & 7.03E-06 & S22E58 & 21 & 461.9  & 34.3/67.5  & Yes& Yes& No  & 120 & 55  \\
72 & 11991 & 2014 Feb 28 & M1.1 & 1.22E-05 & S23E52 & 61 & 333.6  & 123.5/59.6 & Yes& No & No  & -   & -   \\
73 & 11991 & 2014 Mar 05 & C4.8 & 5.04E-06 & S27W07 & 24 & 248.4  &    -       & Yes& No & Yes & -   & -   \\
74 & 11991 & 2014 Mar 05 & C2.8 & 2.90E-06 & S27W08 & 79 & 248.9  &    -       & Yes& No & No  & -   & -   \\
75 & 11991 & 2014 Mar 05 & M1.0 & 1.06E-05 & S27W08 & 7  & 318.7  & 27.8/89.8  & Yes& No & No  & -   & -   \\
76 & 12017 & 2014 Mar 28 & M2.0 & 2.06E-05 & N12W21 & 42 & 2256.6 &    -       & Yes& Yes& Yes & 260 & 31  \\
77 & 12017 & 2014 Mar 28 & M2.6 & 2.67E-05 & N12W24 & 37 & 2150.1 &    -       & Yes& Yes& Yes & 514 & 138 \\
78 & 12017 & 2014 Mar 29 & X1.0 & 1.02E-04 & N11W33 & 64 & 2030.6 & 28.8/65.7  & Yes& No & Yes & 528 & 360 \\
79 & 12031 & 2014 Apr 06 & C3.8 & 3.96E-06 & N03W23 & 66 & 1476.1 &    -       & No & No & Yes & -   & -   \\
80 & 12035 & 2014 Apr 15 & C8.6 & 9.08E-06 & S15E25 & 21 & 1030.0 & 12.5/93.6  & Yes& Yes& Yes & 274 & 27  \\
81 & 12035 & 2014 Apr 15 & C7.3 & 7.92E-06 & S14E21 & 47 & 2640.7 &    -       & No & Yes& Yes & 360 & 179 \\
82 & 12035 & 2014 Apr 16 & M1.0 & 1.04E-05 & S13E08 & 33 & 1920.4 & 150.4/148.0& Yes& Yes& Yes & 764 & 61  \\
83 & 12036 & 2014 Apr 18 & M7.3 & 7.32E-05 & S17W29 & 138& 23961.9&    -       & No & Yes& No  & 1203& 360 \\
84 & 12087 & 2014 Jun 13 & M2.6 & 2.66E-05 & S20E40 & 71 & 1004.1 & 33.0/91.5  & Yes& Yes& No  & 370 & 42  \\
85 & 12087 & 2014 Jun 13 & C9.0 & 9.17E-06 & S19E32 & 26 & 909.3  & 16.0/79.3  & No & Yes& No  & 605 & 31  \\
86 & 12127 & 2014 Jul 31 & C1.3 & 1.44E-06 & S07E32 & 8  & 193.0  &    -       & Yes& Yes& Yes & 458 & 77  \\
87 & 12146 & 2014 Aug 22 & C2.2 & 2.22E-06 & N11E01 & 88 & 3099.4 &    -       & No & Yes& No  & -   & -   \\
88 & 12148 & 2014 Aug 22 & C6.6 & 6.71E-06 & N08W32 & 27 & 394.4  & 7.2/87.4   & Yes& No & No  & -   & -   \\
89 & 12157 & 2014 Sep 13 & C3.7 & 3.92E-06 & S16W39 & 34 & 315.3  & 17.5/35.3  & No & Yes& No  & -   & -   \\
90 & 12192 & 2014 Oct 20 & M1.4 & 1.57E-05 & S15E46 & 21 & 508.6  &    -       & Yes& Yes& No  & -   & -   \\
91 & 12201 & 2014 Nov 03 & C4.2 & 4.32E-06 & S03E21 & 26 & 987.7  &    -       & Yes& Yes& Yes & -   & -   \\
92 & 12227 & 2014 Dec 13 & C4.0 & 4.10E-06 & S03W66 & 36 & 1040.3 &    -       & Yes& Yes& Yes & 198 & 74  \\
93 & 12242 & 2014 Dec 17 & M8.7 & 8.76E-05 & S22E09 & 162& 13165.7& 358.8/186.2& No & No & No  & 587 & 360 \\
94 & 12266 & 2015 Jan 19 & C3.3 & 3.41E-06 & S06E04 & 23 & 415.0  &    -       & Yes& No & Yes & -   & -   \\
95 & 12268 & 2015 Jan 29 & C8.2 & 8.32E-06 & S12W03 & 119& 2071.2 & 26.9/129.6 & No & No & No  & -   & -   \\
96 & 12268 & 2015 Jan 29 & M2.1 & 2.12E-05 & S11W07 & 99 & 3042.0 & 30.9/116.0 & No & Yes& No  & -   & -   \\
97 & 12268 & 2015 Jan 29 & C6.4 & 6.43E-06 & S11W11 & 191& 3235.1 & 31.2/112.5 & No & No & No  & -   & -   \\
98 & 12268 & 2015 Jan 30 & M2.0 & 2.09E-05 & S12W15 & 205& 4132.4 & 37.3/104.7 & No & No & No  & -   & -   \\
99 & 12268 & 2015 Jan 30 & M1.7 & 1.77E-05 & S11W17 & 116& 4783.7 & 87.8/87.2  & No & No & No  & -   & -   \\
100 & 12276 & 2015 Jan 30 & C3.8 & 3.84E-06 & S07E09 & 88 & 957.5  &    -       & Yes& Yes& Yes & -   & -   \\
101 & 12277 & 2015 Feb 03 & C3.9 & 3.98E-06 & N07W04 & 43 & 599.0  & 14.2/77.4  & Yes& Yes& Yes & -   & -   \\
102 & 12297 & 2015 Mar 09 & C9.1 & 9.40E-06 & S15E44 & 33 & 771.9  & 34.5/52.8  & Yes& Yes& Yes & 583 & 155 \\
103 & 12297 & 2015 Mar 10 & M5.1 & 5.29E-05 & S14E39 & 25 & 1182.6 &    -       & Yes& Yes& No  & 1040& 360 \\
104 & 12297 & 2015 Mar 10 & C1.3 & 4.67E-06 & S16E41 & 38 & 574.7  &    -       & Yes& Yes& No  & -   & -   \\
105 & 12297 & 2015 Mar 11 & M2.9 & 2.97E-05 & S15E27 & 85 & 591.8  &    -       & No & Yes& No  & 702 & 160 \\
106 & 12297 & 2015 Mar 12 & M2.7 & 2.81E-05 & S14E02 & 94 & 828.0  &    -       & No & Yes& No  & -   & -   \\
107 & 12297 & 2015 Mar 13 & M1.8 & 1.89E-05 & S14W03 & 80 & 710.5  &    -       & Yes& Yes& No  & -   & -   \\
108 & 12297 & 2015 Mar 15 & C2.4 & 2.51E-06 & S17W28 & 16 & 554.1  &    -       & Yes& No & Yes & -   & -   \\
109 & 12297 & 2015 Mar 15 & C1.0 & 1.04E-06 & S17W40 & 21 & 346.6  &    -       & Yes& No & No  & -   & -   \\
110 & 12325 & 2015 Apr 16 & C3.3 & 3.37E-06 & N06E51 & 112& 1177.4 & 12.8/47.5  & No & No & No  & -   & -   \\
111 & 12434 & 2015 Oct 15 & C3.6 & 3.77E-06 & S11E55 & 32 & 417.2  &    -       & No & No & No  & -   & -   \\
112 & 12434 & 2015 Oct 15 & C3.9 & 4.07E-06 & S11E53 & 34 & 627.7  &    -       & No & No & No  & -   & -   \\
113 & 12434 & 2015 Oct 15 & C3.4 & 3.51E-06 & S11E52 & 49 & 535.3  &    -       & No & No & No  & -   & -   \\
114 & 12434 & 2015 Oct 15 & C3.1 & 3.22E-06 & S12E51 & 47 & 440.8  & 29.5/92.2  & No & No & No  & 283 & 13  \\
115 & 12434 & 2015 Oct 15 & M1.1 & 1.21E-05 & S13E50 & 13 & 561.7  & 375.9/178.2 & No & No & No  & -   & -   \\
116 & 12434 & 2015 Oct 16 & C3.4 & 3.58E-06 & S13E45 & 16 & 569.5  & 95.7/167.6 & Yes& Yes& No  & -   & -   \\
117 & 12434 & 2015 Oct 16 & C3.1 & 3.22E-06 & S13E44 & 23 & 563.5  & 281.9/177.0 & No & Yes& No  & -   & -   \\
118 & 12434 & 2015 Oct 16 & M1.1 & 1.14E-05 & S13E46 & 26 & 599.0  & 282.8/177.8 & No & Yes& Yes & -   & -   \\
119 & 12434 & 2015 Oct 16 & C4.2 & 4.34E-06 & S13E42 & 14 & 608.8  & 262.2/170.7 & Yes & Yes& Yes & 189 & 83  \\
120 & 12434 & 2015 Oct 24 & C1.3 & 1.35E-06 & S13W73 & 29 & 720.4  &    -       & Yes & No & No  & -   & -   \\
121 & 12497 & 2016 Feb 13 & C1.3 & 1.37E-06 & N14W25 & 13 & 241.5  & 29.2/89.2  & No & No & No  & -   & -   \\
122 & 12497 & 2016 Feb 13 & C2.8 & 2.85E-06 & N14W26 & 27 & 281.5  & 25.1/83.1  & Yes & No & No  & 278 & 19  \\
123 & 12497 & 2016 Feb 13 & B8.5 & 8.68E-07 & N14W26 & 26 & 130.3  &    -       & No & No & No  & -   & -   \\
124 & 12497 & 2016 Feb 13 & B6.9 & 7.03E-07 & N14W27 & 51 & 123.1  &    -       & No & No & No  & -   & -   \\
125 & 12497 & 2016 Feb 13 & M1.8 & 1.88E-05 & N14W29 & 120& 361.1  & 59.2/72.8  & Yes & No & No  & -   & -   \\
126 & 12497 & 2016 Feb 13 & C1.6 & 1.67E-06 & N14W31 & 26 & 400.8  & 49.0/66.4  & No & No & No  & -   & -   \\
127 & 12497 & 2016 Feb 14 & C3.4 & 3.48E-06 & N14W36 & 70 & 464.8  & 51.7/50.7  & No & No & No  & -   & -   \\
128 & 12497 & 2016 Feb 14 & M1.0 & 1.05E-05 & N14W48 & 39 & 987.2  & 52.7/85.5  & No & No & No  & -   & -   \\
129 & 12497 & 2016 Feb 15 & C3.9 & 3.94E-06 & N14W53 & 31 & 899.1  & 26.2/28.6  & Yes& No & No  & -   & -   \\
130 & 12567 & 2016 Jul 16 & C6.8 & 7.11E-06 & N05E26 & 35 & 260.6  & 19.4/31.4  & No & No & No  & -   & -   \\
131 & 12615 & 2016 Nov 30 & C2.3 & 3.90E-06 & S07E43 & 24 & 127.3  &    -       & No & No & No  & -   & -   \\
132 & 12661 & 2017 Jun 03 & C2.1 & 2.16E-06 & N07E56 & 77 & 636.9  &    -       & No & Yes & Yes & -   & -   \\
133 & 12661 & 2017 Jun 03 & C2.5 & 2.67E-06 & N06E53 & 46 & 921.5  &    -       & No & Yes & No  & -   & -   \\
134 & 12661 & 2017 Jun 05 & B5.8 & 5.90E-07 & N06E18 & 45 & 155.5  & 6.2/60.5   & No & No & No  & -   & -   \\
\hline
\end{longtable*}
\end{center}

\section{Statistical Properties} \label{sta}

\begin{figure}
\includegraphics[width=0.45\textwidth,clip=]{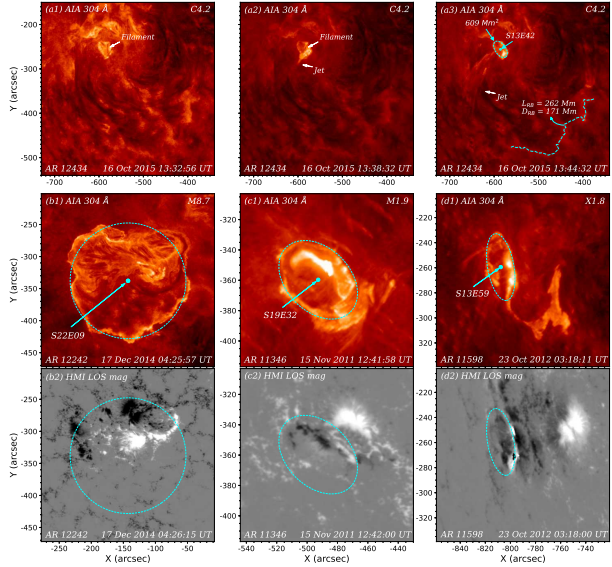}
\centering
\caption{Top panels: Three snapshots of the AIA 304 {\AA} images, showing the evolution of the C4.2 class flare in AR 12434 on 2015 October 16.
The white arrows point to the accompanying mini-filament eruption and coronal jet.
In panel (a3), the location and total area of the flare are labeled. The remote brightening is outlined by the cyan line. 
The total length and average distance between the flare and remote brightening are labeled.
Middle panels: Additional three cases observed in 304 {\AA}. The flare ribbons are fitted with a circle and two ellipses (dashed cyan lines). The centers are pointed by solid arrows.
Bottom panels: Corresponding LOS magnetograms of the three events observed by HMI.}
\label{fig1}
\end{figure}

\subsection{Location} \label{loc}
\begin{figure}
\includegraphics[width=0.45\textwidth,clip=]{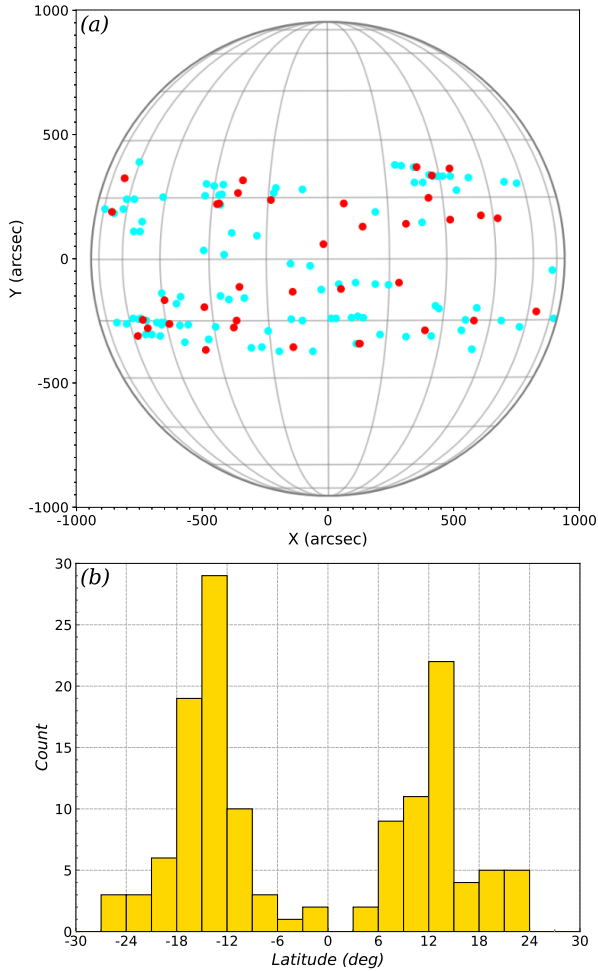}
\centering
\caption{(a) Locations of the 134 CFs, which are marked with cyan dots (confined flares) and red dots (eruptive flares), respectively.
The latitudes and longitudes of the Sun are drawn with gray lines.
(b) Latitude distribution of the CFs. Positive (negative) latitude means northern (southern) hemisphere, respectively.}
\label{fig2}
\end{figure}

In Figure~\ref{fig1}, the top panels show three images of the C4.2 class flare observed by AIA in 304 {\AA} on 2015 October 16 \citep{Zha2016a,Dai2021}.
The flare was triggered by a mini-filament eruption and was associated with a blowout coronal jet.
In panel (a3), the flare ribbon is fitted with an ellipse, whose center is considered as the flare location (S13E42).
The middle panels of Figure~\ref{fig1} show 304 {\AA} images of another three cases. The bright ribbons are fitted with a circle and two ellipses, which are drawn with dashed cyan lines.
The centers (S22E09, S19E32, and S13E59) of the flares are derived and pointed by solid arrows.
The bottom panels of Figure~\ref{fig1} show the corresponding LOS magnetograms of the three events observed by HMI.
The CFs are characterized by a central positive (negative) polarity surrounded by negative (positive) polarities, 
which is essentially consistent with the fan-spine magnetic structure \citep{Mas2009,Wang2012}. Most of the CFs in our sample show similar characteristics.
For all 134 events, we fit the circular or quasi-circular ribbons observed in 304 or 1600 {\AA} with a circle or an ellipse. 
The derived coordinates of CFs in the sixth column of Table~\ref{tab:list} agree with those recorded in the GOES flare catalog \footnote{https://hesperia.gsfc.nasa.gov/goes/goes\_event\_listings/}.
In Figure~\ref{fig2}(a), the flare centers are marked with colored dots, with cyan (red) dots signifying confined (eruptive) flares, respectively.
Figure~\ref{fig2}(b) shows latitude distribution of the CFs, where 76 (58) of which are located in the southern (northern) hemisphere, respectively.
It is clear that all CFs are located in ARs like microflares, and the latitudes of CFs are between -30$\degr$ and 30$\degr$. Hence, the distribution of CFs is consistent with the AR belts.

\begin{deluxetable*}{lccccccc}
\tablecaption{Minima, maxima, mean values, and median values of the physical properties of CFs and the related activities. 
\label{tab:prop}}
\tablecolumns{8}
\tablenum{2}
\tablewidth{0pt}
\tablehead{
\colhead{parameter} &
\colhead{$A_{CF}$} &
\colhead{$r_{CF}$} & 
\colhead{$\tau_{CF}$} &
\colhead{$L_{RB}$} &
\colhead{$D_{RB}$} &
\colhead{$V_{CME}$} &
\colhead{$W_{CME}$} \\
\colhead{} &
\colhead{(Mm$^2$)} &
\colhead{(Mm)} &
\colhead{(minute)} &
\colhead{(Mm)} &
\colhead{(Mm)} &
\colhead{(km s$^{-1}$)} &
\colhead{($\degr$)}
}
\startdata
Minimum & 123.1     & 6.3  & 4.0   & 6.2   & 28.6  & 120.0  & 13.0  \\
Maximum & 23961.9   & 87.3 & 205.0 & 381.0 & 186.2 & 1203.0 & 360.0 \\
Mean & 1214.8 & 16.8 & 49.6  & 84.1  & 89.1 & 516.7  & 143.6 \\
Median & 631.2 & 14.2  & 35.0 & 44.3 &  84.8  & 514.0  &  80.0 \\
\enddata
\end{deluxetable*}

\subsection{Area} \label{area}
\begin{figure}
\includegraphics[width=0.45\textwidth,clip=]{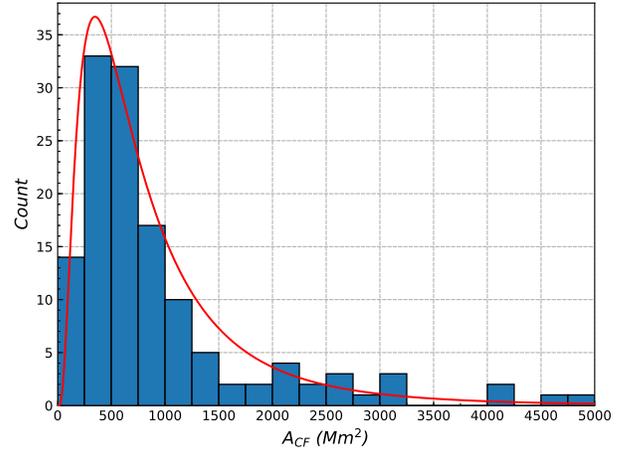}
\centering
\caption{Distribution of the flare area. The result of curve-fitting using the log-normal function is superposed with a red line.
Two values significantly exceeding 5000 Mm$^{2}$ are excluded in this plot.}
\label{fig3}
\end{figure}

Compared with two-ribbon flares whose ribbons show separation \citep{qiu02} or elongation \citep{Li2015,Qiu2017}, the outer ribbons of CFs hardly expand. 
Hence, the area keeps constant. In Figure~\ref{fig1}(a3), the apparent area of the C4.2 flare enclosed by the ellipse is $\sim$452 Mm$^{2}$.
The true area ($\sim$609 Mm$^{2}$) after deprojection according to the flare longitude is defined as the total area ($A_{CF}$).
In the middle panels of Figure~\ref{fig1}, the apparent areas of the three CFs are $\sim$13008, $\sim$985 and $\sim$739 Mm$^{2}$.
The true areas are estimated to be $\sim$13166, $\sim$1165 and $\sim$1423 Mm$^{2}$, respectively.
For the 134 CFs, the areas are calculated according to the fitted circles or ellipses, which range from 123 to $\sim$5000 Mm$^{2}$ in most cases.
Figure~\ref{fig3} shows the distribution of $A_{CF}$. Note that the areas (13166 and 23962 Mm$^{2}$) of two events significantly exceeding 5000 Mm$^{2}$ are not taken into account in the histogram. 
We fit the distribution of $A_{CF}$ with a log-normal function \citep{Zha2010}:
\begin{equation} \label{eqn-1}
  \frac{1}{N}\frac{dN}{dx}=f(x,\mu,\sigma)=\frac{1}{x\sigma\sqrt{2\pi}}e^{-\frac{(\ln x-\mu)^2}{2\sigma^2}}, x>0.
\end{equation}
The curve fitting is performed by using \texttt{mpfit.pro} and the result is superposed with a red line in Figure~\ref{fig3}, where $\mu=6.51$ and $\sigma=0.66$.

\begin{figure}
\includegraphics[width=0.45\textwidth,clip=]{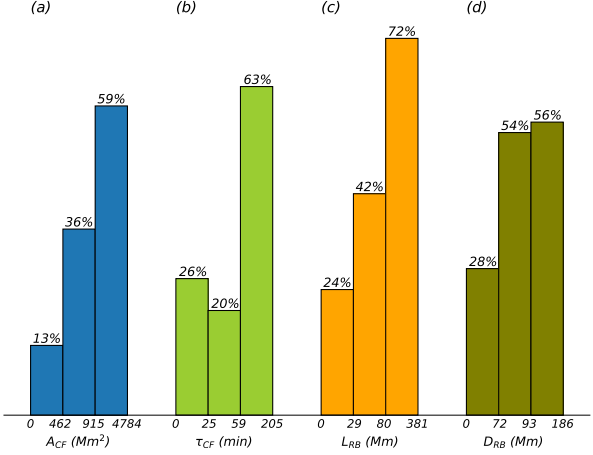}
\centering
\caption{Percentages of M- and X-class CFs in different groups of area (a), lifetime (b), total length of remote brightening (c), and average distance of remote brightening (d).}
\label{fig4}
\end{figure}

We notice that there is a tendency of increasing area with flare class. In Figure~\ref{fig4}(a), the flare areas are divided into three groups of roughly the same amount.
In the range of 0$-$462  Mm$^{2}$, only 13\% (6/45) are M- and X-class CFs. The proportion rises to 36\% (16/45) in the range of 462$-$915 Mm$^{2}$. 
Above 915 Mm$^{2}$, M- and X-class CFs account for 59\% (26/44). Therefore, flares with larger magnitudes generally have larger areas.
The sample of CFs in our study overlaps with that in \citet{ST2018}. They used an irregularly closed line rather than a circle or an oval to fit the flare ribbon.
Although the methods are different, the derived areas of CFs are very close to each other.

It is widely believed that the magnetic configuration of CFs consist of a null point and the associated dome-shaped fan-spine structure. 
Assuming that the fan surface is semi-sphere \citep{Par2009,Par2010}, the height of null point is equivalent to the radius of circular ribbons:
\begin{equation} \label{eqn-2}
  h_{NP}=r_{CF}=\sqrt{\frac{A_{CF}}{\pi}}.
\end{equation}

\begin{figure}
\includegraphics[width=0.45\textwidth,clip=]{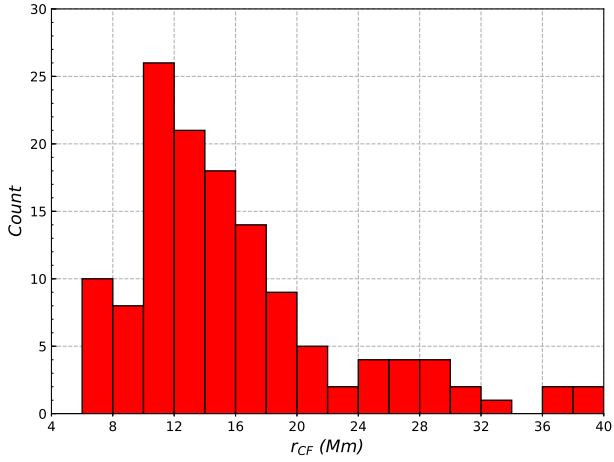}
\centering
\caption{Distribution of the equivalent radius of flare ribbons.}
\label{fig5}
\end{figure}

Figure~\ref{fig5} shows the distribution of $r_{CF}$, which lies in the range of 6$-$39 Mm with a mean value of $\sim$16 Mm (the largest two events are excluded).
The estimated altitudes of null points are consistent with previous results \citep[e.g.,][]{Sun2013,Xu2017,Hou2019,LY2019,Yang2020b}.
\citet{ST2018} found that WL flares feature smaller areas than normal flares with the same flare magnitude,
which is explained by the lower heights of null points where magnetic free energy is released and consequently larger energy fluxes in WL flares.

\subsection{Lifetime} \label{life}
\begin{figure}
\includegraphics[width=0.45\textwidth,clip=]{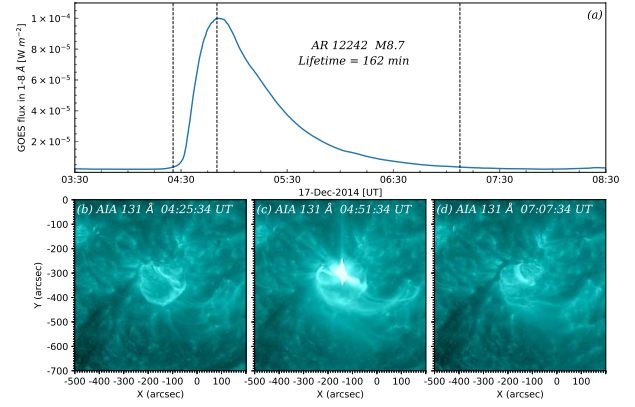}
\centering
\caption{(a) SXR light curve of the M8.7 class flare occurring in AR 12242 on 2014 December 17.
The three black dashed lines denote the start, peak, and end times of the flare.
(b-d) AIA 131 {\AA} images of the flare.}
\label{fig6}
\end{figure}

We determine the flare lifetime ($\tau_{CF}$) using the start and end times.
The former refers to the time when the GOES 1$-$8 {\AA} flux begins to increase rapidly, and the latter refers to the time when the flux declines to a nearly constant level.
In Figure~\ref{fig6}, the bottom panels show AIA 131 {\AA} images of the M8.7 class flare occurring in AR 12242 on 2014 December 17 \citep{Chen2019}. 
The top panel shows the SXR light curve of the flare. The black dashed lines denote the start time (04:25:34 UT), peak time (04:51:34 UT), and end time (07:07:34 UT) of the flare. 
The lifetime reaches $\sim$162 minutes accordingly.

\begin{figure}
\includegraphics[width=0.45\textwidth,clip=]{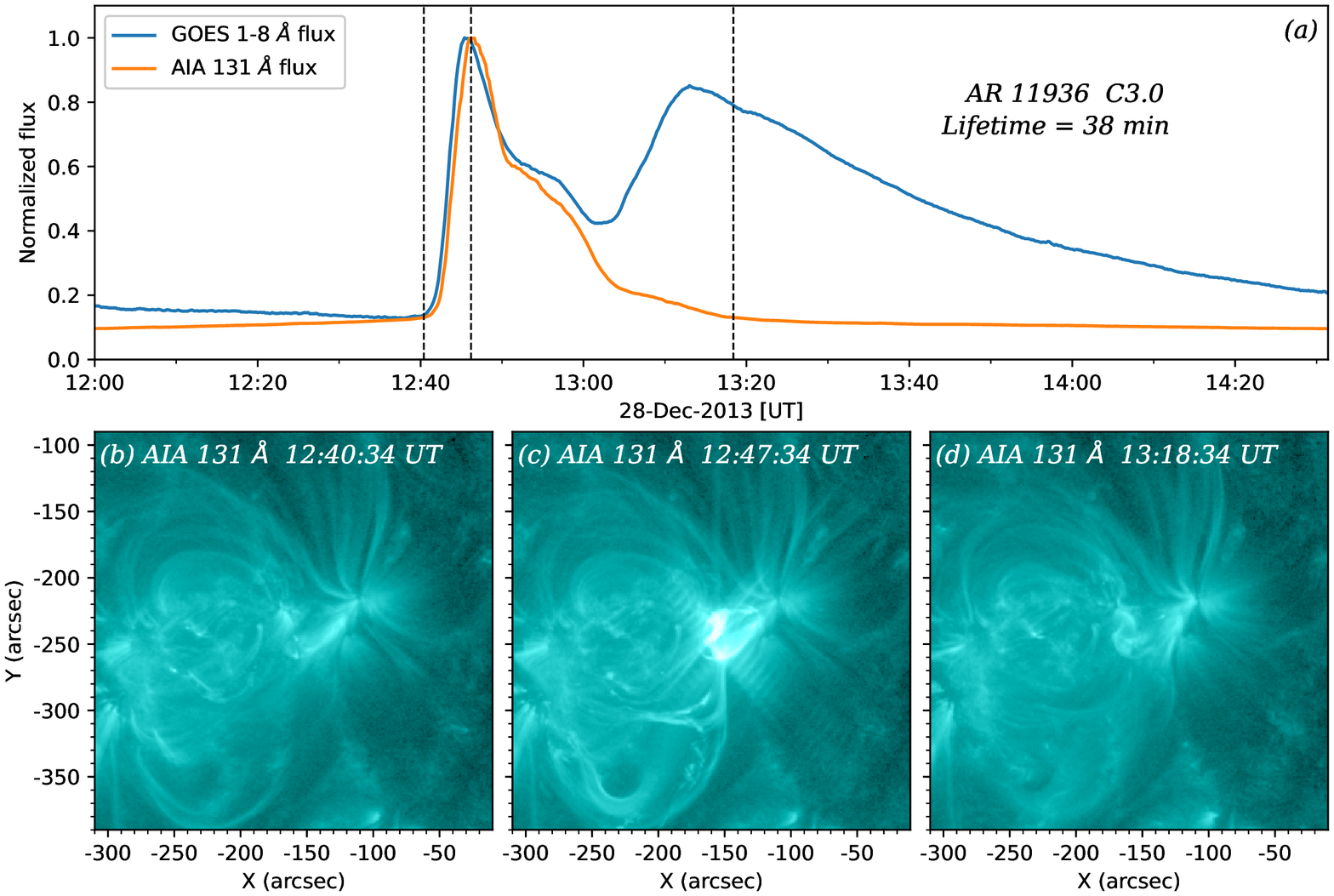}
\centering
\caption{(a) SXR light curve (blue line) and 131 {\AA} light curve (orange line) of the C3.0 class flare occurring in AR 11936 on 2013 December 28.
The three black dashed lines denote the start, peak, and end times of the flare.
(b-d) AIA 131 {\AA} images of the flare.}
\label{fig7}
\end{figure}

If the SXR light curve has a second peak in the decay phase, which is probably due to another eruption somewhere else, 
the AIA 131 {\AA} light curve of the relevant flare is used as a supplementary to determine $\tau_{CF}$.
In Figure~\ref{fig7}, the bottom panels show AIA 131 {\AA} images of the C3.0 class flare occurring in AR 11936 on 2013 December 28.
The top panel shows the SXR light curve (blue line) and 131 {\AA} light curve (orange line) of the flare, respectively. 
The black dashed lines denote the start time (12:40:34 UT), first peak time (12:47:34 UT), and end time (13:18:34 UT).
There is only one peak in the EUV light curve, which corresponds to the first peak in SXR. The second peak ($\sim$13:14:55 UT) in SXR is unrelated to the C3.0 class flare.
Hence, the lifetime of flare is $\sim$38 minutes, which is considerably shorter than the M8.7 flare.

\begin{figure}
\includegraphics[width=0.45\textwidth,clip=]{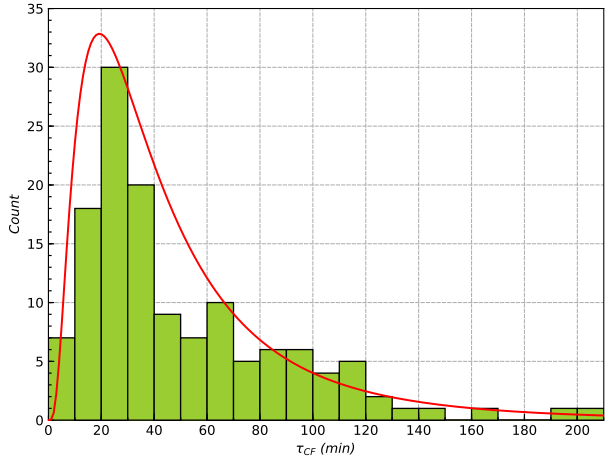}
\centering
\caption{Distribution of the lifetime of CFs. The result of curve-fitting using the log-normal function is superposed with a red line.}
\label{fig8}
\end{figure}

\begin{figure}
\includegraphics[width=0.45\textwidth,clip=]{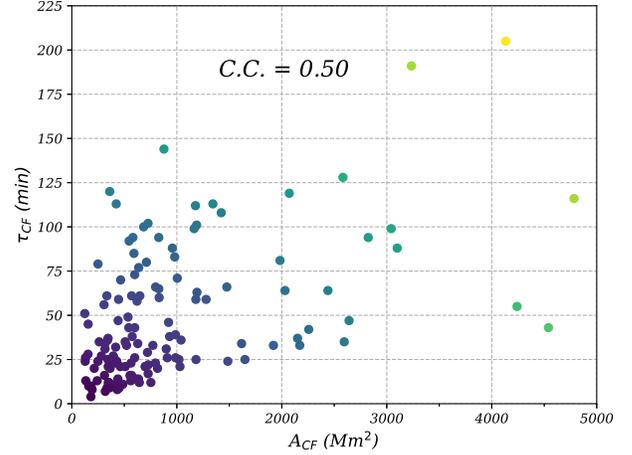}
\centering
\caption{Scatter plot between the area ($A_{CF}$) and lifetime ($\tau_{CF}$) of CFs.
The correlation coefficient is $\sim$0.50.}
\label{fig9}
\end{figure}

Figure~\ref{fig8} shows the distribution of $\tau_{CF}$ with a mean value of $\sim$50 minutes, which is two times larger than that of Ly$\alpha$ flares \citep{Lu2021}.
Likewise, the distribution can be fitted with a log-normal function in Equation~\ref{eqn-1}. The fitted curve is superposed with a red line, where $\mu=3.60$ and $\sigma=0.64$.
To explore the relationship between area and lifetime of CFs, we draw a scatter plot in Figure~\ref{fig9}. The two parameters have a positive correlation with a coefficient of $\sim$0.5.
The timescale of conductive cooling of hot flare loops is expressed as \citep{Car1994,Zha2019a}:
\begin{equation} \label{eqn-3}
  \tau_{cc}=4\times10^{-10}\frac{n_{\mathrm{e}}L^2}{T_{\mathrm{e}}^{5/2}},
\end{equation}
where $n_\mathrm{e}$, $T_\mathrm{e}$, and $L$ represent the electron number density, temperature, and total length of a flare loop.
For CFs with fan-spine topology, $A_{CF}\propto L^2\propto \tau_{cc}$. 
That is to say, for flares with larger sizes, the cooling times become longer, which can easily interpret the linear correlation between $A_{CF}$ and $\tau_{CF}$.

According to the lifetime, we divide the 134 flares into three groups of the same amount, i.e., 0$-$25, 25$-$59, and 59$-$205 minutes.
The percentages of M- and X-class CFs in the three groups are 26\% (12/46), 20\% (9/45), and 63\% (27/43) (see Figure~\ref{fig4}(b)).
Hence, flares with larger magnitudes tend to have longer lifetimes.
In the statistical investigation of CFs \citep{ST2018}, the end time corresponds to the time when the flux decreases to a point halfway between the peak flux and the pre-flare level, 
leading to an average lifetime of 10$-$20 minutes, which is systematically shorter than our results.
Besides, some of the CFs may experience an EUV late phase after the main flare phase observed in ``warm'' emission lines (e.g., 335 {\AA}) as a result of additional heating \citep{Sun2013}.
The much extended lifetimes in EUV 335 {\AA} compared with SXR lifetimes are out of the scope of this study.

\subsection{Peak SXR flux in 1$-$8 {\AA}} \label{peak}
\begin{figure}
\includegraphics[width=0.45\textwidth,clip=]{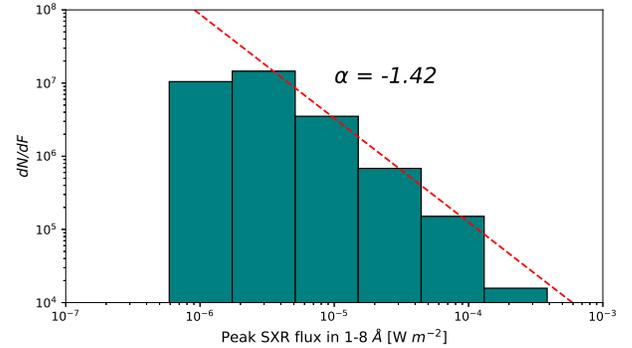}
\centering
\caption{Distribution of the peak SXR flux in 1$-$8 {\AA}. The power-law index $\alpha$ is labeled.}
\label{fig10}
\end{figure}

For the 134 CFs, the peak SXR flux in 1$-$8 {\AA} ranges from $\sim$5.9$\times$10$^{-7}$ to $\sim$3.9$\times$10$^{-4}$ W m$^{-2}$. Figure~\ref{fig10} shows the distribution of the peak SXR flux. 
It is clear that the distribution could be nicely fitted with a power-law function \citep[e.g.,][]{Den1985,Lu1991,Cro1993,Chr2008} above $\sim$2$\times$10$^{-6}$ W m$^{-2}$:
\begin{equation} \label{eqn-4}
  \frac{dN}{dF}\propto F^{\alpha},
\end{equation}
where $F$ denotes the peak flux, and $\alpha=-1.42$ denotes the power-law index. 
The value of $\alpha$ is close to that of microflares at 3$-$6 keV and less than that of Ly$\alpha$ flares \citep{Lu2021}.

\begin{deluxetable}{lccccccc}
\tablecaption{Numbers of CFs associated with RBs, type III radio bursts, jets, mini-filament eruptions, and CMEs. \label{tab:num}}
\tablecolumns{6}
\tablenum{3}
\tablewidth{0pt}
\tablehead{
\colhead{activity} &
\colhead{B-class} &
\colhead{C-class} & 
\colhead{M-class} &
\colhead{X-class} &
\colhead{total} 
}
\startdata
CF  & 4 & 82 & 40 & 8 & 134 \\
RB & 1 & 40  & 28   & 7  &  76 \\
type III & 0 & 36  & 24  & 3  &  63 \\
jet & 1 & 43  & 21   & 4  & 69  \\
FE & 1 & 25  & 18   & 7  &  51 \\
CME & 0 & 17  & 14   & 6  &  37 \\
\enddata
\end{deluxetable}

\subsection{Relation with remote brightenings} \label{rb}
As mentioned in Section~\ref{intro}, CFs are usually associated with remote brightenings or remote ribbons \citep{Mas2009,Xu2017,Song2018,Chen2019}.
In Figure~\ref{fig1}(a3), remote brightenings appear to the southwest of the C4.2 class flare, which is overlaid with a cyan line.
The brightening has a total length ($L_{RB}$) of $\sim$262 Mm and an average distance ($D_{RB}$) of $\sim$171 Mm away from the flare center.

\begin{figure}
\includegraphics[width=0.45\textwidth,clip=]{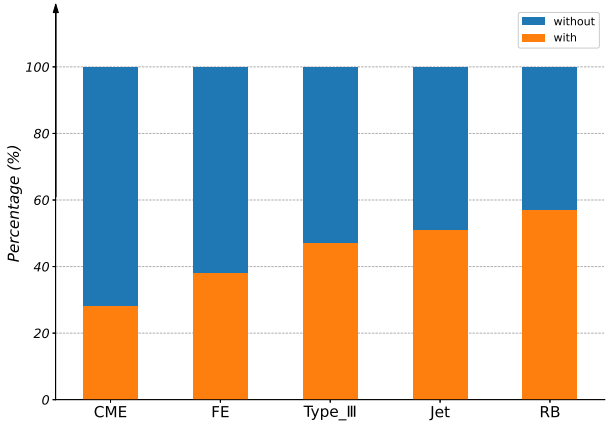}
\centering
\caption{Proportions of CFs related to remote brightenings, type III radio bursts, jets, mini-filament eruptions (FEs), and CMEs.}
\label{fig11}
\end{figure}

\begin{figure}
\includegraphics[width=0.45\textwidth,clip=]{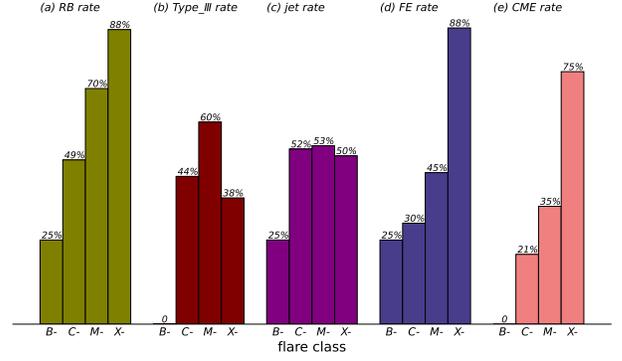}
\centering
\caption{Association rates of CFs with remote brightenings (a), type III radio bursts (b), jets (c), mini-filament eruptions (d), and CMEs (e) for 
B-, C-, M-, and X-class flares, respectively.}
\label{fig12}
\end{figure}

\begin{figure}
\includegraphics[width=0.45\textwidth,clip=]{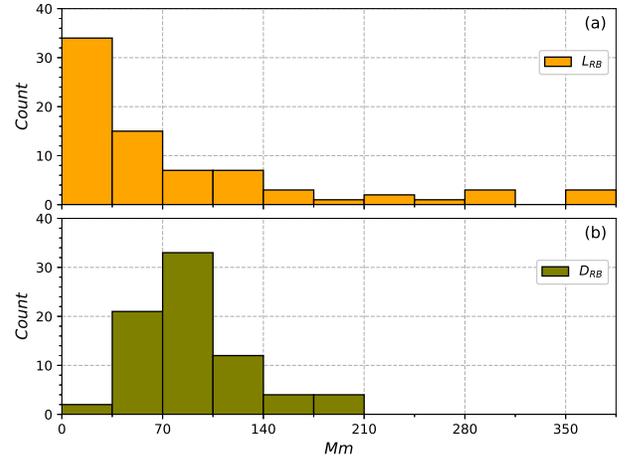}
\centering
\caption{Distributions of the total length ($L_{RB}$) and average distance ($D_{RB}$) of remote brightenings associated with CFs.}
\label{fig13}
\end{figure}

\begin{figure}
\includegraphics[width=0.45\textwidth,clip=]{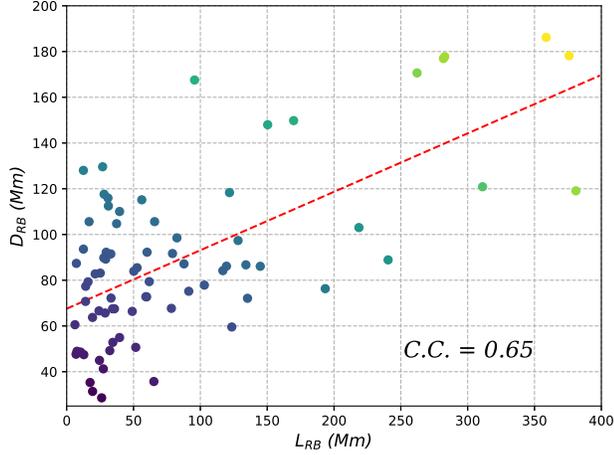}
\centering
\caption{Scatter plot between the total length ($L_{RB}$) and average distance ($D_{RB}$) of remote brightenings. 
The result of linear fitting is plotted with a red dashed line. The correlation coefficient ($\sim$0.65) is labeled.}
\label{fig14}
\end{figure}

For the 134 CFs in this study, we searched for remote brightenings observed by AIA mainly in 304 and 1600 {\AA}, 
finding that $\sim$57\% (76/134) of CFs have associated remote brightenings, which is shown in Figure~\ref{fig11}.
Specifically, the association rates with B-, C-, M-, and X-class flares are 25\%, 49\%, 70\%, 88\%, respectively (see Figure~\ref{fig12}(a) and Table~\ref{tab:num}).
The increasing rate with flare magnitude indicates that larger flares are more likely to produce remote brightenings.
The total length of remote brightening or ribbon is denoted with $L_{RB}$, and the average distance from the flare center is denoted with $D_{RB}$.
Figure~\ref{fig13} shows the distributions of $L_{RB}$ and $D_{RB}$. It is revealed that $L_{RB}$ lies in the range of 6.2$-$381.0 Mm, with a mean value of $\sim$84.1 Mm.
$D_{RB}$ lies in the range of 28.6$-$186.2 Mm, with a mean value of $\sim$89.1 Mm.
Figure~\ref{fig14} shows the scatter plot between $L_{RB}$ and $D_{RB}$, indicating a good correlation between the two parameters with a correlation coefficient of $\sim$0.65.
In other words, remote brightenings further away from the main CFs are more likely to have longer extensions. 
A linear fitting (red dashed line) between the two parameters is performed, i.e., 
\begin{equation} \label{eqn-5}
  D_{RB}=67.56+0.26L_{RB}.
\end{equation}

To investigate the relationship between flare magnitude and association with remote brightenings, we divide $L_{RB}$ and $D_{RB}$ into three groups with the same numbers.
The values of $L_{RB}$ are divided into 0$-$29, 29$-$80, and 80$-$381 Mm. The values of $D_{RB}$ are divided into 0$-$72, 72$-$93, and 93$-$186 Mm.
The percentages of M- and X-class CFs in each group are plotted in Figure~\ref{fig4}(c-d).
The percentages increase systematically with both parameters, indicating that flares with larger magnitudes are more likely to produce longer remote brightenings or ribbons.
In the 3D numerical simulations of jets confined by large-scale coronal loops \citep{WD2016,Wyp2016}, the aspect ratio of the fan-spine structure is between 1.0 and 2.7.
In our study, the aspect ratio is defined as $\frac{D_{RB}}{2r_{CF}}$, which has a range of 0.8$-$6.7 and a mean value of $\sim$2.9.
Hence, the results will impose constraints on numerical simulations of CFs in the future \citep{Pon2013}.

\subsection{Relation with type III radio bursts} \label{rabt}
\begin{figure}
\includegraphics[width=0.45\textwidth,clip=]{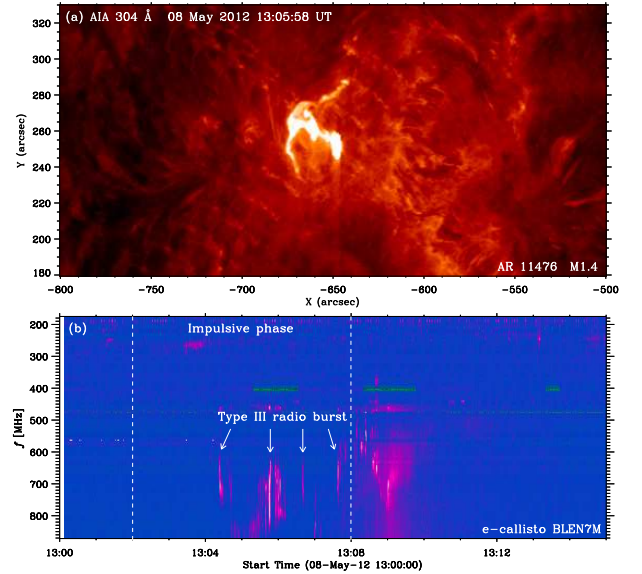}
\centering
\caption{(a) AIA 304 {\AA} image of the M1.4 class CF occurring in AR 11476 on 2012 May 8.
(b) Radio dynamic spectra of the flare recorded by the e-Callisto/BLEN7M station.
The arrows point to the type III radio bursts. The white dashed lines mark the start and peak times of the flare.}
\label{fig15}
\end{figure}

Flare-accelerated nonthermal electrons propagating along open field lines are capable of producing type III radio bursts \citep{Ben2005,Mor2014,RR2014,Zha2015}.
In Figure~\ref{fig15}, the top panel shows the 304 {\AA} image of an M1.4 class CF in AR 11476 on 2012 May 8.
The corresponding radio dynamic spectra recorded by the e-Callisto/BLEN7M station is displayed in the bottom panel.
It is obvious that type III radio bursts around 700 MHz with fast frequency drift occur during the impulsive phase, when the release rate of magnetic free energy is maximum.
The existence of radio bursts is confirmed by the WIND/WAVES observation.

For the 134 CFs, 47\% (63/134) of them are accompanied with type III bursts during the impulsive phases, which is displayed in Figure~\ref{fig11}.
Note that those occurring in the pre-flare or decay phases are considered as unrelated events, which may result in an underestimate of the total amount.
In Figure~\ref{fig12}(b), the proportions of CFs related to type III bursts at various flare magnitudes are demonstrated, being 0\%, 44\%, 60\%, and 38\% 
for B-, C-, M-, and X-class CFs, respectively. The relatively lower proportion in B-class flares may be due to the limited number of sample.
Hence, there is no correlation between the flare magnitude and type III bursts.
In other words, the production of type III bursts depends mainly on the magnetic topology instead of flare energy \citep{Kru2011,Gle2012,Duan2022}.

\subsection{Relation with coronal jets} \label{jet}
As mentioned in Section~\ref{intro}, CFs are frequently associated with hot coronal jets observed in EUV wavelengths \citep{LY2019,ZN2019,ZQM2020,Zha2021} 
or cool surges observed in H$\alpha$ \citep{Wang2012,Xu2017}.
In Figure~\ref{fig1}, the C4.2 class flare is accompanied by a blowout jet \citep{Zha2016a,Dai2021}.
For the 134 events, about half of them are accompanied by jets in 304 {\AA}, which is displayed in Figure~\ref{fig11}.
In Figure~\ref{fig12}(c), the proportions of CFs related to jets at various flare magnitudes are demonstrated, being 25\%, 52\%, 53\%, and 50\% 
for B-, C-, M-, and X-class flares, respectively. Therefore, there is no preference of jet production for CFs with larger magnitudes as well \citep{Chae1999,Shi2007,Duan2022}.

\subsection{Relation with mini-filament eruptions} \label{fila}
The occurrence of CFs is often associated with filament or minifilament eruptions \citep{Wang2012,Jia2013,Sun2013,Yang2018}.
In this scenario, breakout-type magnetic reconnection takes place near the null point after the filament embedded in the fan dome rises up \citep{Jos2015,Wyp2017}.
The initiation of filament eruption may result from magnetic flux emergence \citep{Li2017}, rotation \citep{Xu2017}, or ideal MHD instabilities.
The null-point reconnection not only accelerates nonthermal electrons to produce the circular ribbon in the chromosphere, but speeds up the filament eruption as well \citep{Zha2021}.

Most of the homologous CFs in AR 12434 are triggered by mini-filament eruptions \citep{Zha2016a,Zha2021}.
Figure~\ref{fig1}(a2) shows the AIA 304 {\AA} image of the C4.2 class flare, and the mini-filament within the circular ribbon is pointed by the arrow.
For the 134 events, 38\% (51/134) of them are associated with filament eruptions, which is displayed in Figure~\ref{fig11}.
Figure~\ref{fig12}(d) shows the proportions of CFs related to mini-filament eruptions: 25\% (1/4), 30\% (25/82), 45\% (18/40) and 88\% (7/8) for B-, C-, M-, and X-class flares, respectively.
The increasing proportions with flare classes indicate that larger flares are more likely to be triggered by filament eruptions.

\subsection{Relation with CMEs} \label{cme}
\begin{figure}
\includegraphics[width=0.45\textwidth,clip=]{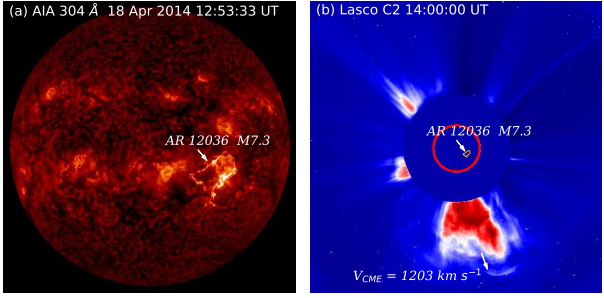}
\centering
\caption{(a) AIA 304 {\AA} image of the M7.3 class flare in AR 12036 on 2014 April 18.
(b) WL image of the associated CME at a speed of 1203 km s$^{-1}$ in the LASCO/C2 FOV.}
\label{fig16}
\end{figure}

As mentioned in the previous section, CFs are sometimes triggered by filament or minifilament eruptions. 
The eruption may also drive a CME observed in the WL coronagraphs \citep{Jos2017,Liu2020,Kum2021}.
In Figure~\ref{fig16}, the left panel shows the 304 {\AA} image of the M7.3 class CF in AR 12036 on 2014 April 18 \citep{Jos2015}.
The associated CME at a speed of $\sim$1203 km s$^{-1}$ in the LASCO/C2 FOV is displayed in the right panel.
For the 134 CFs, only 28\% (37/134) of them are related to CMEs, which is shown in Figure~\ref{fig11}.
It is evident that the association rate with CMEs is the lowest while the association rate with remote brightenings is the highest in our investigation.
In other words, most of ($\sim$72\%) CFs are confined rather than eruptive \citep{WD2016,Li2018,Yang2018}.
The CME association rates for B-, C-, M-, X-class flares are 0\%, 21\%, 35\%, and 75\%, which is displayed in Figure~\ref{fig12}(e).
The increasing CME rates with flare magnitudes are consistent with previous findings, suggesting that larger flares are more likely to produce CMEs \citep{Yas2005,Yas2006}.
The low percentage of CMEs is expected, since magnetic flux ropes are required to generate CMEs in most cases \citep{Vour2013}.
For eruptive flares, two parallel ribbons or \textsf{S}-shaped ribbons appear on both sides of the polarity inversion lines \citep{Aul2012,Jan2013,Sav2016}.   
Therefore, the formation of a CF is unlikely in this situation.

\begin{figure}
\includegraphics[width=0.45\textwidth,clip=]{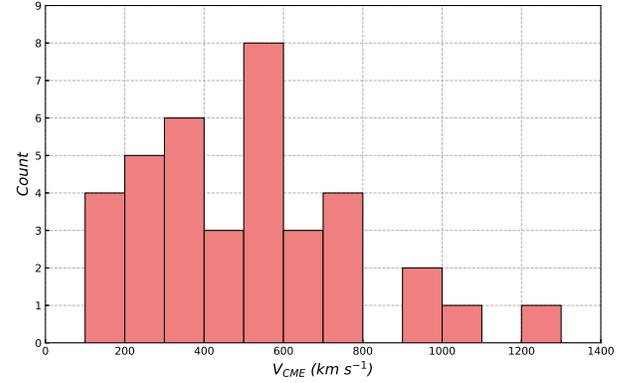}
\centering
\caption{Distribution of the CME speed ($V_{CME}$) related to CFs.}
\label{fig17}
\end{figure}

\begin{figure}
\includegraphics[width=0.45\textwidth,clip=]{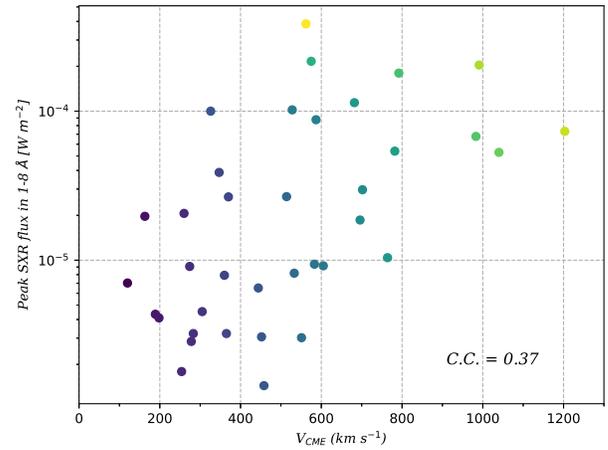}
\centering
\caption{Scatter plot between the peak SXR flux in 1-8 {\AA} of CFs related to CMEs and $V_{CME}$.
The correlation coefficient is $\sim$0.37.}
\label{fig18}
\end{figure}

The distribution of the linear speed ($V_{CME}$) of the 37 CMEs is drawn in Figure~\ref{fig17}. 
The values of $V_{CME}$ lie in the range of 120$-$1203 km s$^{-1}$, with an average speed of $\sim$517 km s$^{-1}$, which is slightly higher than that of CMEs near solar maximum \citep{Yas2004}.
Figure~\ref{fig18} shows the scatter plot between the peak flux in 1$-$8 {\AA} and $V_{CME}$ of the 37 eruptive flares.
A positive correlation is clearly demonstrated with a correlation coefficient of $\sim$0.37, which is consistent with previous finding that CMEs associated with X-class flares are remarkably
faster than those of CMEs associated with C-class flares \citep{Yas2005}.

\begin{figure}
\includegraphics[width=0.45\textwidth,clip=]{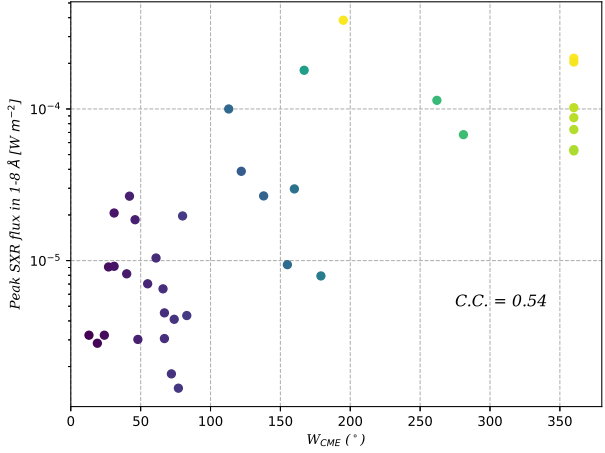}
\centering
\caption{Scatter plot between the peak SXR flux in 1-8 {\AA} of CFs related to CMEs and the CME angular widths ($W_{CME}$).
The correlation coefficient is $\sim$0.54.}
\label{fig19}
\end{figure}

\begin{figure}
\includegraphics[width=0.45\textwidth,clip=]{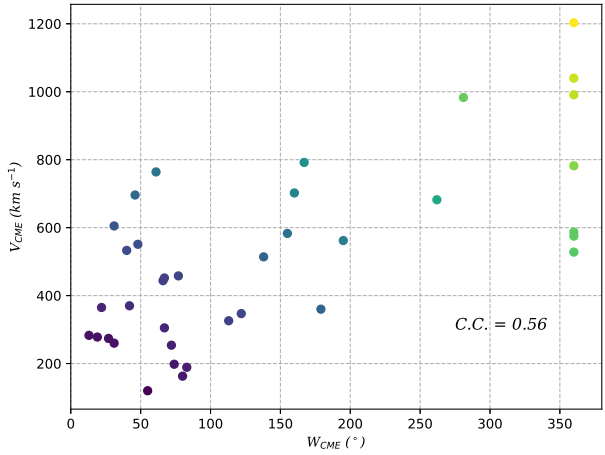}
\centering
\caption{Scatter plot between $V_{CME}$ and $W_{CME}$, indicating faster CMEs tend to be wider.
The correlation coefficient is $\sim$0.56.}
\label{fig20}
\end{figure}

Out of the 37 CMEs, 30 of them are normal or partial halo CMEs, and seven of them are full halo CMEs.
Some of them are jet-like, narrow CMEs with angular width ($W_{CME}$) less than 25$\degr$.
Figure~\ref{fig19} and Figure~\ref{fig20} show scatter plots between $W_{CME}$ and peak flux of flare and $V_{CME}$, respectively.
It is clear that $W_{CME}$ have positive correlations with flare peak flux and $V_{CME}$. 
Note that full halo CMEs ($W_{CME}=360\degr$) are not included when calculating the correlation coefficients.
The result in Figure~\ref{fig20} suggests that faster CMEs tend to be wider, which is in line with previous statistical investigation \citep{Yas2004}.

\begin{figure}
\includegraphics[width=0.45\textwidth,clip=]{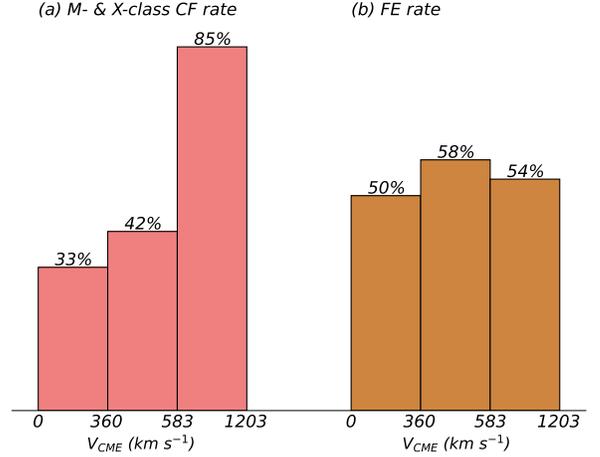}
\centering
\caption{(a) Percentages of M- and X-class flares for the three groups of CMEs.
(b) Percentages of mini-filament eruptions for the three groups of CMEs.}
\label{fig21}
\end{figure}

According to their speeds, the 37 CMEs are divided into three groups with roughly the same numbers.
Figure~\ref{fig21}(a) shows that the percentages of M- and X-class CFs are 33\% (4/12), 42\% (5/12) and 85\% (11/13) for 0$-$360, 360$-$583, and 583$-$1203 km s$^{-1}$, respectively.
The association rates of CMEs with big flares increase with the CME speeds, which is consistent with the fact that faster CMEs are more related to flares with larger magnitudes \citep{Yas2005,Yas2006}.
Figure~\ref{fig21}(b) shows the percentages of filament eruptions for the three groups of CMEs, being 50\% (6/12), 58\% (7/12) and 54\% (7/13).
The association rates of CMEs with filament eruptions are independent of the CME speeds.

\begin{figure}
\includegraphics[width=0.45\textwidth,clip=]{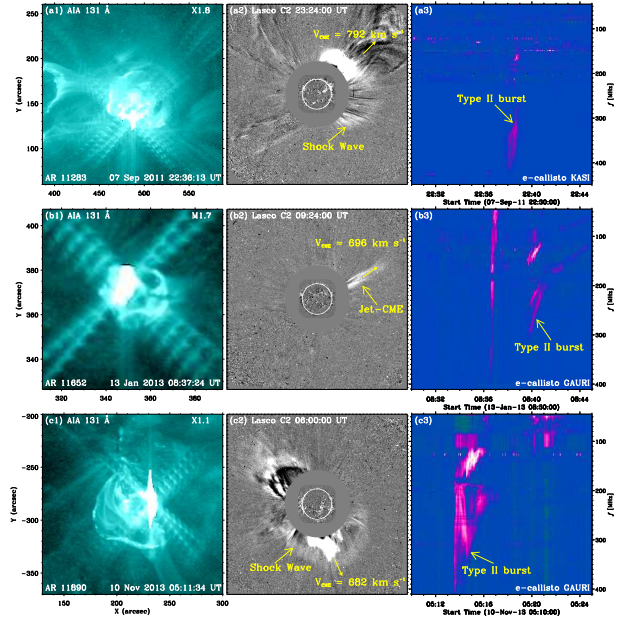}
\centering
\caption{Left panels: Three CFs observed by AIA in 131 {\AA}.
Middle panels: Associated CMEs and shock waves observed by LASCO/C2. The apparent speeds of the CMEs are labeled.
Right panels: Radio dynamic spectra of the three events recorded by the e-Callisto stations, showing the corresponding type II radio bursts with drifting frequency.}
\label{fig22}
\end{figure}

Fast CMEs are capable of driving shock waves, which are associated with type II radio bursts \citep[e.g.,][]{On2009,Zu2018,Man2019}.
For the 37 eruptive CFs accompanied by CMEs, seven of which are associated with type II bursts, including two X-class (No. 3, 53) and five M-class (No. 34, 49, 77, 82, 83) flares.
In Figure~\ref{fig22}, the left panels show three CFs observed in 131 {\AA}. The middle panels show the corresponding CMEs observed by LASCO/C2, with the apparent speeds being labeled.
The related shock waves with relatively lower intensities are pointed by the arrows. The right panels show radio dynamic spectra recorded by the e-Callisto stations, 
where type II radio bursts with drifting frequency are pointed by the arrows.

\section{Conclusion and discussion} \label{sum}
In this paper, we conducted a comprehensive statistical analysis of 134 CFs observed by SDO/AIA from 2011 September to 2017 June, 
including four B-class, 82 C-class, 40 M-class, and eight X-class flares, respectively.
The physical properties of CFs are derived, including the location, area ($A_{CF}$), equivalent radius ($r_{CF}$) assuming a semi-spherical fan dome, 
lifetime ($\tau_{CF}$), and peak SXR flux in 1$-$8 {\AA}.
Combining with the observations from the SOHO/LASCO, WIND/WAVES, and radio stations of the e-Callisto network,
we also explored the relations with remote brightenings, type III radio bursts, coronal jets, mini-filament eruptions, and CMEs.
The main results are as follows:

\begin{enumerate}
\item{All CFs are located in active regions, with the latitudes between -30$\degr$ and 30$\degr$.
Most of the areas are $\leq$5000 Mm$^2$, with a mean value of $\sim$1215 Mm$^2$.
The distribution could be fitted with a log-normal function, where $\mu=6.51$, and $\sigma=0.66$.
The equivalent radii, representing the heights of magnetic null points, are between 6 and 87 Mm, with a mean value of $\sim$16.8 Mm.}
\item{The lifetimes are between 4 and 205 minutes with a mean value of $\sim$50 minutes.
The distribution could also be fitted with a log-normal function, where $\mu=3.6$, and $\sigma=0.64$.
There is a positive correlation between the lifetime and area, indicating that CFs with larger sizes tend to have longer lifetimes.
The peak SXR flux in 1$-$8 {\AA} is well in accord with a power-law distribution with an index of $-$1.42.}
\item{For the 134 CFs, 57\% of them are accompanied by remote brightenings or ribbons. The association rates with RBs increase with flare magnitudes.
The mean values of total length ($L_{RB}$) and average distance ($D_{RB}$) of remote brightenings are $\sim$84 Mm and $\sim$89 Mm, respectively.
A positive correlation exists between the two parameters, i.e., $D_{RB}=67.56+0.26L_{RB}$.}
\item{About 47\% and 51\% of the 134 CFs are related to type III radio bursts and jets, respectively. The association rates are independent of flare classes, 
meaning that the production rates of type III radio bursts and jets depend mainly on the magnetic configuration instead of the flare energy.
About 38\% of CFs are related to mini-filament eruptions, and the association rates increase with flare classes.}
\item{Only 28\% of CFs are related to CMEs, meaning that a majority of them are confined rather than eruptive events.
This is in agreement with the relatively high association with remote brighenings considering the particular magnetic topology of CFs.
The association rates with CMEs increase with flare magnitudes. The apparent speeds of CMEs are between 120 and 1203 km s$^{-1}$, with a mean value of $\sim$517 km s$^{-1}$.
There is a positive correlation between the CME speed and peak SXR flux in 1$-$8 {\AA}. The angular widths of CMEs ($W_{CME}$) are between 13$\degr$ and 360$\degr$.
A linear correlation exists between the angular width and apparent speed, indicating that faster CMEs tend to be wider.}
\end{enumerate}

\citet{Li2021} analyzed 719 flares during 2010$-$2019, finding that the total unsigned magnetic flux ($\Phi_{AR}$) plays a decisive role in determining the eruptive character of a flare.
For the 134 CFs, only 37 events are eruptive. The mean and median values of total area for the eruptive CFs are significantly greater than confined CFs.
Hence, it is worthwhile investigating the total magnetic fluxes of CFs and justifying the conclusion of \citet{Li2021}.

Of course, the current statistical work has limitations. Firstly, the sample is not large enough, leading to uncertainties in the results.
Those flares with irregular or complex ribbons are excluded to minimize the ambiguity. 
The 134 flares were observed by SDO/AIA from 2011 September to 2017 June, including four B-class and eight X-class flares.
The numbers of CFs will certainly grow after extending the dates to 2021 June for instance.
Secondly, the shapes of outer ribbons are fitted with circles or ellipses, which is suitable in most cases. However, the fitting is unsatisfactory for irregular ribbons \citep{Jos2015}.
The method in \citet{ST2018} seems to be more appropriate. Besides, the assumption of semi-spherical fan surface is not always valid to estimate the heights of null points \citep{VW2014}.
Thirdly, the values of apparent speeds and angular widths of CMEs suffer from projection effect from a single viewpoint.
Three-dimensional reconstruction of the shapes of CMEs is helpful to minimize this effect \citep{Zha2022}.
Finally, the properties of CFs are still incomplete. The distribution of thermal energy and association rate with QPPs will be the focus of our next paper.
The results are important to get a better understanding of solar flares and provide valuable constraints on 3D MHD simulations.
In the future, new insights into CFs and the related activities will be obtained from observations of the Spectrometer/Telescope for Imaging X-rays \citep[STIX;][]{Kru2020} 
on board Solar Orbiter \citep{Mu2020}.

\begin{acknowledgments}
The authors appreciate the referee for valuable suggestions and comments. The e-Callisto data are courtesy of the Institute for Data Science FHNW Brugg/Windisch, Switzerland.
SDO is a mission of NASA\rq{}s Living With a Star Program. AIA and HMI data are courtesy of the NASA/SDO science teams.
This work is supported by the NSFC grants (No. 11790302, 11790300), and the Strategic Priority Research Program on Space Science, CAS (XDA15052200, XDA15320301).
\end{acknowledgments}


\begin{thebibliography}{}
\bibitem[Aulanier et al.(2007)]{Aul2007} Aulanier, G., Golub, L., DeLuca, E.~E., et al.\ 2007, Science, 318, 1588. doi:10.1126/science.1146143
\bibitem[Aulanier et al.(2012)]{Aul2012} Aulanier, G., Janvier, M., \& Schmieder, B.\ 2012, \aap, 543, A110. doi:10.1051/0004-6361/201219311
\bibitem[Benz et al.(2005)]{Ben2005} Benz, A.~O., Grigis, P.~C., Csillaghy, A., et al.\ 2005, \solphys, 226, 121. doi:10.1007/s11207-005-5254-5
\bibitem[Bougeret et al.(1995)]{Bou1995} Bougeret, J.-L., Kaiser, M.~L., Kellogg, P.~J., et al.\ 1995, \ssr, 71, 231. doi:10.1007/BF00751331
\bibitem[Brueckner et al.(1995)]{Bru1995} Brueckner, G.~E., Howard, R.~A., Koomen, M.~J., et al.\ 1995, \solphys, 162, 357. doi:10.1007/BF00733434
\bibitem[Cai et al.(2021)]{Cai2021} Cai, Z.~M., Zhang, Q.~M., Ning, Z.~J., et al.\ 2021, \solphys, 296, 61. doi:10.1007/s11207-021-01805-5
\bibitem[Cargill(1994)]{Car1994} Cargill, P.~J.\ 1994, \apj, 422, 381. doi:10.1086/173733
\bibitem[Carmichael(1964)]{Car1964} Carmichael, H.\ 1964, NASA Special Publication, 451
\bibitem[Carrington(1859)]{Car1859} Carrington, R.~C.\ 1859, \mnras, 20, 13. doi:10.1093/mnras/20.1.13
\bibitem[Chae et al.(1999)]{Chae1999} Chae, J., Qiu, J., Wang, H., et al.\ 1999, \apjl, 513, L75. doi:10.1086/311910
\bibitem[Chen et al.(2019)]{Chen2019} Chen, X., Yan, Y., Tan, B., et al.\ 2019, \apj, 878, 78. doi:10.3847/1538-4357/ab1d64
\bibitem[Christe et al.(2008)]{Chr2008} Christe, S., Hannah, I.~G., Krucker, S., et al.\ 2008, \apj, 677, 1385. doi:10.1086/529011
\bibitem[Crosby et al.(1993)]{Cro1993} Crosby, N.~B., Aschwanden, M.~J., \& Dennis, B.~R.\ 1993, \solphys, 143, 275. doi:10.1007/BF00646488
\bibitem[Dai et al.(2021)]{Dai2021} Dai, J., Zhang, Q.~M., Su, Y.~N., et al.\ 2021, \aap, 646, A12. doi:10.1051/0004-6361/202039013
\bibitem[Demoulin et al.(1996)]{Dem1996} Demoulin, P., Henoux, J.~C., Priest, E.~R., et al.\ 1996, \aap, 308, 643
\bibitem[Dennis(1985)]{Den1985} Dennis, B.~R.\ 1985, \solphys, 100, 465. doi:10.1007/BF00158441
\bibitem[De Pontieu et al.(2014)]{De2014} De Pontieu, B., Title, A.~M., Lemen, J.~R., et al.\ 2014, \solphys, 289, 2733. doi:10.1007/s11207-014-0485-y
\bibitem[Devi et al.(2020)]{Devi2020} Devi, P., Joshi, B., Chandra, R., et al.\ 2020, \solphys, 295, 75. doi:10.1007/s11207-020-01642-y
\bibitem[Duan et al.(2022)]{Duan2022} Duan, Y., Shen, Y., Zhou, X., et al.\ 2022, \apjl, 926, L39. doi:10.3847/2041-8213/ac4df2
\bibitem[Fletcher et al.(2011)]{Fle2011} Fletcher, L., Dennis, B.~R., Hudson, H.~S., et al.\ 2011, \ssr, 159, 19. doi:10.1007/s11214-010-9701-8
\bibitem[Glesener et al.(2012)]{Gle2012} Glesener, L., Krucker, S., \& Lin, R.~P.\ 2012, \apj, 754, 9. doi:10.1088/0004-637X/754/1/9
\bibitem[Hannah et al.(2008)]{Han2008} Hannah, I.~G., Christe, S., Krucker, S., et al.\ 2008, \apj, 677, 704. doi:10.1086/529012
\bibitem[Hao et al.(2017)]{Hao2017} Hao, Q., Yang, K., Cheng, X., et al.\ 2017, Nature Communications, 8, 2202. doi:10.1038/s41467-017-02343-0
\bibitem[Hernandez-Perez et al.(2017)]{Her2017} Hernandez-Perez, A., Thalmann, J.~K., Veronig, A.~M., et al.\ 2017, \apj, 847, 124. doi:10.3847/1538-4357/aa8814
\bibitem[Hirayama(1974)]{Hir1974} Hirayama, T.\ 1974, \solphys, 34, 323. doi:10.1007/BF00153671
\bibitem[Hou et al.(2019)]{Hou2019} Hou, Y., Li, T., Yang, S., et al.\ 2019, \apj, 871, 4. doi:10.3847/1538-4357/aaf4f4
\bibitem[Janvier et al.(2013)]{Jan2013} Janvier, M., Aulanier, G., Pariat, E., et al.\ 2013, \aap, 555, A77. doi:10.1051/0004-6361/201321164
\bibitem[Ji et al.(2006)]{ji06} Ji, H., Huang, G., Wang, H., et al.\ 2006, \apjl, 636, L173. doi:10.1086/500203
\bibitem[Jiang et al.(2013)]{Jia2013} Jiang, C., Feng, X., Wu, S.~T., et al.\ 2013, \apjl, 771, L30. doi:10.1088/2041-8205/771/2/L30
\bibitem[Jing et al.(2016)]{Jing2016} Jing, J., Xu, Y., Cao, W., et al.\ 2016, Scientific Reports, 6, 24319. doi:10.1038/srep24319
\bibitem[Joshi et al.(2015)]{Jos2015} Joshi, N.~C., Liu, C., Sun, X., et al.\ 2015, \apj, 812, 50. doi:10.1088/0004-637X/812/1/50
\bibitem[Joshi et al.(2017)]{Jos2017} Joshi, N.~C., Sterling, A.~C., Moore, R.~L., et al.\ 2017, \apj, 845, 26. doi:10.3847/1538-4357/aa7c1b
\bibitem[Joshi et al.(2018)]{Jos2018} Joshi, N.~C., Nishizuka, N., Filippov, B., et al.\ 2018, \mnras, 476, 1286. doi:10.1093/mnras/sty322
\bibitem[Joshi et al.(2021)]{Jos2021} Joshi, N.~C., Joshi, B., \& Mitra, P.~K.\ 2021, \mnras, 501, 4703. doi:10.1093/mnras/staa3480
\bibitem[Kashapova et al.(2020)]{Kash2020} Kashapova, L.~K., Kupriyanova, E.~G., Xu, Z., et al.\ 2020, \aap, 642, A195. doi:10.1051/0004-6361/201833947
\bibitem[Kopp \& Pneuman(1976)]{Kop1976} Kopp, R.~A. \& Pneuman, G.~W.\ 1976, \solphys, 50, 85.doi:10.1007/BF00206193
\bibitem[Krucker et al.(2011)]{Kru2011} Krucker, S., Kontar, E.~P., Christe, S., et al.\ 2011, \apj, 742, 82. doi:10.1088/0004-637X/742/2/82
\bibitem[Krucker et al.(2020)]{Kru2020} Krucker, S., Hurford, G.~J., Grimm, O., et al.\ 2020, \aap, 642, A15. doi:10.1051/0004-6361/201937362
\bibitem[Kumar et al.(2015)]{Kum2015} Kumar, P., Nakariakov, V.~M., \& Cho, K.-S.\ 2015, \apj, 804, 4. doi:10.1088/0004-637X/804/1/4
\bibitem[Kumar et al.(2016)]{Kum2016} Kumar, P., Nakariakov, V.~M., \& Cho, K.-S.\ 2016, \apj, 822, 7. doi:10.3847/0004-637X/822/1/7
\bibitem[Kumar et al.(2021)]{Kum2021} Kumar, P., Karpen, J.~T., Antiochos, S.~K., et al.\ 2021, \apj, 907, 41. doi:10.3847/1538-4357/abca8b
\bibitem[Lee et al.(2020a)]{Lee2020a} Lee, J., Karpen, J.~T., Liu, C., et al.\ 2020, \apj, 893, 158. doi:10.3847/1538-4357/ab80c4
\bibitem[Lee et al.(2020b)]{Lee2020b} Lee, J., White, S.~M., Chen, X., et al.\ 2020, \apjl, 901, L10. doi:10.3847/2041-8213/abb4dd
\bibitem[Lemen et al.(2012)]{Lem2012} Lemen, J.~R., Title, A.~M., Akin, D.~J., et al.\ 2012, \solphys, 275, 17. doi:10.1007/s11207-011-9776-8
\bibitem[Li et al.(2015)]{Li2015} Li, D., Ning, Z.~J., \& Zhang, Q.~M.\ 2015, \apj, 807, 72. doi:10.1088/0004-637X/807/1/72
\bibitem[Li et al.(2017)]{Li2017} Li, H., Jiang, Y., Yang, J., et al.\ 2017, \apj, 836, 235. doi:10.3847/1538-4357/aa5eac
\bibitem[Li et al.(2018)]{Li2018} Li, T., Yang, S., Zhang, Q., et al.\ 2018, \apj, 859, 122. doi:10.3847/1538-4357/aabe84
\bibitem[Li \& Yang(2019)]{LY2019} Li, H. \& Yang, J.\ 2019, \apj, 872, 87. doi:10.3847/1538-4357/aafb3a
\bibitem[Li et al.(2020)]{Li2020} Li, D., Feng, S., Su, W., et al.\ 2020, \aap, 639, L5. doi:10.1051/0004-6361/202038398
\bibitem[Li et al.(2021)]{Li2021} Li, T., Chen, A., Hou, Y., et al.\ 2021, \apjl, 917, L29. doi:10.3847/2041-8213/ac1a15
\bibitem[Liu et al.(2013)]{Liu2013} Liu, C., Deng, N., Lee, J., et al.\ 2013, \apjl, 778, L36. doi:10.1088/2041-8205/778/2/L36
\bibitem[Liu et al.(2015)]{Liu2015} Liu, C., Deng, N., Liu, R., et al.\ 2015, \apjl, 812, L19. doi:10.1088/2041-8205/812/2/L19
\bibitem[Liu et al.(2019)]{Liu2019} Liu, C., Lee, J., \& Wang, H.\ 2019, \apj, 883, 47. doi:10.3847/1538-4357/ab3923
\bibitem[Liu et al.(2020)]{Liu2020} Liu, C., Prasad, A., Lee, J., et al.\ 2020, \apj, 899, 34. doi:10.3847/1538-4357/ab9cbe
\bibitem[Lu \& Hamilton(1991)]{Lu1991} Lu, E.~T. \& Hamilton, R.~J.\ 1991, \apjl, 380, L89. doi:10.1086/186180
\bibitem[Lu et al.(2021)]{Lu2021} Lu, L., Feng, L., Li, D., et al.\ 2021, \apjs, 253, 29. doi:10.3847/1538-4365/abd79b
\bibitem[Mancuso et al.(2019)]{Man2019} Mancuso, S., Frassati, F., Bemporad, A., et al.\ 2019, \aap, 624, L2. doi:10.1051/0004-6361/201935157
\bibitem[Masson et al.(2009)]{Mas2009} Masson, S., Pariat, E., Aulanier, G., et al.\ 2009, \apj, 700, 559. doi:10.1088/0004-637X/700/1/559
\bibitem[Masson et al.(2017)]{Mas2017} Masson, S., Pariat, {\'E}., Valori, G., et al.\ 2017, \aap, 604, A76. doi:10.1051/0004-6361/201629654
\bibitem[Mitra \& Joshi(2021)]{Mi2021} Mitra, P.~K. \& Joshi, B.\ 2021, \mnras, 503, 1017. doi:10.1093/mnras/stab175
\bibitem[Morosan et al.(2014)]{Mor2014} Morosan, D.~E., Gallagher, P.~T., Zucca, P., et al.\ 2014, \aap, 568, A67. doi:10.1051/0004-6361/201423936
\bibitem[M{\"u}ller et al.(2020)]{Mu2020} M{\"u}ller, D., St. Cyr, O.~C., Zouganelis, I., et al.\ 2020, \aap, 642, A1. doi:10.1051/0004-6361/202038467
\bibitem[Ning et al.(2022)]{Ning2022} Ning, Z., Wang, Y., Hong, Z., et al.\ 2022, \solphys, 297, 2. doi:10.1007/s11207-021-01935-w
\bibitem[Ontiveros \& Vourlidas(2009)]{On2009} Ontiveros, V. \& Vourlidas, A.\ 2009, \apj, 693, 267. doi:10.1088/0004-637X/693/1/267
\bibitem[Pariat et al.(2009)]{Par2009} Pariat, E., Antiochos, S.~K., \& DeVore, C.~R.\ 2009, \apj, 691, 61. doi:10.1088/0004-637X/691/1/61
\bibitem[Pariat et al.(2010)]{Par2010} Pariat, E., Antiochos, S.~K., \& DeVore, C.~R.\ 2010, \apj, 714, 1762. doi:10.1088/0004-637X/714/2/1762
\bibitem[Pesnell et al.(2012)]{Pes2012} Pesnell, W.~D., Thompson, B.~J., \& Chamberlin, P.~C.\ 2012, \solphys, 275, 3. doi:10.1007/s11207-011-9841-3
\bibitem[Pontin et al.(2007)]{Pon2007} Pontin, D.~I., Bhattacharjee, A., \& Galsgaard, K.\ 2007, Physics of Plasmas, 14, 052106. doi:10.1063/1.2722300
\bibitem[Pontin et al.(2013)]{Pon2013} Pontin, D.~I., Priest, E.~R., \& Galsgaard, K.\ 2013, \apj, 774, 154. doi:10.1088/0004-637X/774/2/154
\bibitem[Priest \& Pontin(2009)]{Pri2009} Priest, E.~R. \& Pontin, D.~I.\ 2009, Physics of Plasmas, 16, 122101. doi:10.1063/1.3257901
\bibitem[Qiu et al.(2002)]{qiu02} Qiu, J., Lee, J., Gary, D.~E., et al.\ 2002, \apj, 565, 1335. doi:10.1086/324706
\bibitem[Qiu et al.(2017)]{Qiu2017} Qiu, J., Longcope, D.~W., Cassak, P.~A., et al.\ 2017, \apj, 838, 17. doi:10.3847/1538-4357/aa6341
\bibitem[Reid \& Ratcliffe(2014)]{RR2014} Reid, H.~A.~S. \& Ratcliffe, H.\ 2014, Research in Astronomy and Astrophysics, 14, 773. doi:10.1088/1674-4527/14/7/003
\bibitem[Reid et al.(2012)]{Rei2012} Reid, H.~A.~S., Vilmer, N., Aulanier, G., et al.\ 2012, \aap, 547, A52. doi:10.1051/0004-6361/201219562
\bibitem[Savcheva et al.(2016)]{Sav2016} Savcheva, A., Pariat, E., McKillop, S., et al.\ 2016, \apj, 817, 43. doi:10.3847/0004-637X/817/1/43
\bibitem[Scherrer et al.(2012)]{sch12} Scherrer, P.~H., Schou, J., Bush, R.~I., et al.\ 2012, \solphys, 275, 207. doi:10.1007/s11207-011-9834-2
\bibitem[Shen et al.(2019)]{Shen2019} Shen, Y., Qu, Z., Zhou, C., et al.\ 2019, \apjl, 885, L11. doi:10.3847/2041-8213/ab4cf3
\bibitem[Shibata \& Magara(2011)]{Shi2011} Shibata, K. \& Magara, T.\ 2011, Living Reviews in Solar Physics, 8, 6. doi:10.12942/lrsp-2011-6
\bibitem[Shibata et al.(2007)]{Shi2007} Shibata, K., Nakamura, T., Matsumoto, T., et al.\ 2007, Science, 318, 1591. doi:10.1126/science.1146708
\bibitem[Song \& Tian(2018)]{ST2018} Song, Y. \& Tian, H.\ 2018, \apj, 867, 159. doi:10.3847/1538-4357/aae5d1
\bibitem[Song et al.(2018)]{Song2018} Song, Y.~L., Guo, Y., Tian, H., et al.\ 2018, \apj, 854, 64. doi:10.3847/1538-4357/aaa7f1
\bibitem[Sturrock(1966)]{Stu1966} Sturrock, P.~A.\ 1966, \nat, 211, 695. doi:10.1038/211695a0
\bibitem[Sun et al.(2013)]{Sun2013} Sun, X., Hoeksema, J.~T., Liu, Y., et al.\ 2013, \apj, 778, 139. doi:10.1088/0004-637X/778/2/139
\bibitem[Temmer et al.(2001)]{Tem2001} Temmer, M., Veronig, A., Hanslmeier, A., et al.\ 2001, \aap, 375, 1049. doi:10.1051/0004-6361:20010908
\bibitem[Vemareddy \& Wiegelmann(2014)]{VW2014} Vemareddy, P. \& Wiegelmann, T.\ 2014, \apj, 792, 40. doi:10.1088/0004-637X/792/1/40
\bibitem[Veronig et al.(2002)]{Ver2002} Veronig, A., Temmer, M., Hanslmeier, A., et al.\ 2002, \aap, 382, 1070. doi:10.1051/0004-6361:20011694
\bibitem[Vourlidas et al.(2013)]{Vour2013} Vourlidas, A., Lynch, B.~J., Howard, R.~A., et al.\ 2013, \solphys, 284, 179. doi:10.1007/s11207-012-0084-8
\bibitem[Wang \& Liu(2012)]{Wang2012} Wang, H. \& Liu, C.\ 2012, \apj, 760, 101. doi:10.1088/0004-637X/760/2/101
\bibitem[Wyper \& DeVore(2016)]{WD2016} Wyper, P.~F. \& DeVore, C.~R.\ 2016, \apj, 820, 77. doi:10.3847/0004-637X/820/1/77
\bibitem[Wyper et al.(2016)]{Wyp2016} Wyper, P.~F., DeVore, C.~R., Karpen, J.~T., et al.\ 2016, \apj, 827, 4. doi:10.3847/0004-637X/827/1/4
\bibitem[Wyper et al.(2017)]{Wyp2017} Wyper, P.~F., Antiochos, S.~K., \& DeVore, C.~R.\ 2017, \nat, 544, 452. doi:10.1038/nature22050
\bibitem[Xu et al.(2017)]{Xu2017} Xu, Z., Yang, K., Guo, Y., et al.\ 2017, \apj, 851, 30. doi:10.3847/1538-4357/aa9995
\bibitem[Yang et al.(2015)]{Yang2015} Yang, K., Guo, Y., \& Ding, M.~D.\ 2015, \apj, 806, 171. doi:10.1088/0004-637X/806/2/171
\bibitem[Yang \& Zhang(2018)]{Yang2018} Yang, S. \& Zhang, J.\ 2018, \apjl, 860, L25. doi:10.3847/2041-8213/aacaf9
\bibitem[Yang et al.(2020a)]{Yang2020a} Yang, J., Hong, J., Li, H., et al.\ 2020, \apj, 900, 158. doi:10.3847/1538-4357/aba7c0
\bibitem[Yang et al.(2020b)]{Yang2020b} Yang, S., Zhang, Q., Xu, Z., et al.\ 2020, \apj, 898, 101. doi:10.3847/1538-4357/ab9ac7
\bibitem[Yashiro et al.(2004)]{Yas2004} Yashiro, S., Gopalswamy, N., Michalek, G., et al.\ 2004, Journal of Geophysical Research (Space Physics), 109, A07105. doi:10.1029/2003JA010282
\bibitem[Yashiro et al.(2005)]{Yas2005} Yashiro, S., Gopalswamy, N., Akiyama, S., et al.\ 2005, Journal of Geophysical Research (Space Physics), 110, A12S05. doi:10.1029/2005JA011151
\bibitem[Yashiro et al.(2006)]{Yas2006} Yashiro, S., Akiyama, S., Gopalswamy, N., et al.\ 2006, \apjl, 650, L143. doi:10.1086/508876
\bibitem[Yashiro et al.(2008)]{Yas2008} Yashiro, S., Michalek, G., \& Gopalswamy, N.\ 2008, Annales Geophysicae, 26, 3103. doi:10.5194/angeo-26-3103-2008
\bibitem[Zhang(2020)]{ZQM2020} Zhang, Q.~M.\ 2020, \aap, 642, A159. doi:10.1051/0004-6361/202038557
\bibitem[Zhang(2022)]{Zha2022} Zhang, Q.~M.\ 2022, arXiv:2202.10676
\bibitem[Zhang \& Ni(2019)]{ZN2019} Zhang, Q.~M. \& Ni, L.\ 2019, \apj, 870, 113. doi:10.3847/1538-4357/aaf391
\bibitem[Zhang et al.(2010)]{Zha2010} Zhang, Q.-M., Guo, Y., Chen, P.-F., et al.\ 2010, Research in Astronomy and Astrophysics, 10, 461. doi:10.1088/1674-4527/10/5/006
\bibitem[Zhang et al.(2015)]{Zha2015} Zhang, Q.~M., Ning, Z.~J., Guo, Y., et al.\ 2015, \apj, 805, 4. doi:10.1088/0004-637X/805/1/4
\bibitem[Zhang et al.(2016a)]{Zha2016a} Zhang, Q.~M., Li, D., Ning, Z.~J., et al.\ 2016a, \apj, 827, 27. doi:10.3847/0004-637X/827/1/27
\bibitem[Zhang et al.(2016b)]{Zha2016b} Zhang, Q.~M., Li, D., \& Ning, Z.~J.\ 2016b, \apj, 832, 65. doi:10.3847/0004-637X/832/1/65
\bibitem[Zhang et al.(2019a)]{Zha2019a} Zhang, Q.~M., Li, D., \& Huang, Y.\ 2019a, \apj, 870, 109. doi:10.3847/1538-4357/aaf4b7
\bibitem[Zhang et al.(2019b)]{Zha2019b} Zhang, Q.~M., Cheng, J.~X., Feng, L., et al.\ 2019b, \apj, 883, 124. doi:10.3847/1538-4357/ab3a52
\bibitem[Zhang et al.(2020)]{Zha2020} Zhang, Q.~M., Yang, S.~H., Li, T., et al.\ 2020, \aap, 636, L11. doi:10.1051/0004-6361/202038072
\bibitem[Zhang et al.(2021)]{Zha2021} Zhang, Q.~M., Huang, Z.~H., Hou, Y.~J., et al.\ 2021, \aap, 647, A113. doi:10.1051/0004-6361/202038924
\bibitem[Zimovets et al.(2021)]{Zim2021} Zimovets, I.~V., McLaughlin, J.~A., Srivastava, A.~K., et al.\ 2021, \ssr, 217, 66. doi:10.1007/s11214-021-00840-9
\bibitem[Zucca et al.(2018)]{Zu2018} Zucca, P., Morosan, D.~E., Rouillard, A.~P., et al.\ 2018, \aap, 615, A89. doi:10.1051/0004-6361/201732308
\end{thebibliography}

\end{document}